\documentclass[useAMS,usenatbib]{mn2e}
\pdfoutput=1
\usepackage[pdftex]{graphicx}
\usepackage{amsmath}
\usepackage[utf8]{inputenc}
\usepackage{pdfpages}
\usepackage{aas_macros}
\long\def\symbolfootnote[#1]#2{\begingroup%
\def\thefootnote{\fnsymbol{footnote}}\footnote[#1]{#2}\endgroup}

\title[RR Lyrae in the Hercules-Aquila Cloud]
  {Strong RR Lyrae excess in the Hercules-Aquila Cloud}
\author[I.T. Simion et al.]
  {Iulia T.~Simion$^1$\thanks{email: isimion@ast.cam.ac.uk},
  Vasily~Belokurov$^1$, Mike~Irwin$^1$, Sergey E.~Koposov$^{1,2}$
\\ $^{1}$Institute of Astronomy, Madingley Rd, Cambridge, CB3 0HA,
\\ $^{2}$Moscow MV Lomonosov State University, Sternberg Astronomical 
Institute, Moscow 119992, Russia
}
\date{Accepted 2014 January 16.  Received 2014 January 15; in original form 2013 December 16}

\pagerange{\pageref{firstpage}--\pageref{lastpage}} \pubyear{2014}

\def\LaTeX{L\kern-.36em\raise.3ex\hbox{a}\kern-.15em
    T\kern-.1667em\lower.7ex\hbox{E}\kern-.125emX}

\begin{document}

\label{firstpage}

\maketitle

\begin{abstract}
We map the large-scale sub-structure in the Galactic stellar halo
using accurate 3D positions of $\sim$14,000 RR Lyrae reported by the
Catalina Sky Survey. In the heliocentric distance range of 10-25 kpc,
in the region of the sky approximately bounded by $30^{\circ} < l <
55^{\circ}$ and $-45^{\circ} < b < -25^{\circ}$, there appears to be a
strong excess of RRab stars. This overdensity, peaking at 18 kpc, is
most likely associated with the so-called Hercules-Aquila Cloud,
previously detected using Main Sequence tracers at similar distances
in the Sloan Digital Sky Survey data. Our analysis of the
period-amplitude distribution of RR Lyrae in this region indicates
that the HAC is dominated by the Oosterhoff I type population. By
comparing the measured RR Lyrae number density to models of a smooth
stellar halo, we estimate the significance of the observed excess and
provide an updated estimate of the total luminosity of the Cloud's
progenitor.
\end{abstract}



\section{Introduction}

With the recent discoveries of both ongoing disruption events
\citep[such as the Sagittarius stream, see e.g.][]{Ma2003} and of the
remnants of past mergers \citep[e.g. the Virgo Cloud,
  see][]{Duffau2006,Ju2008}, there is now little doubt that a
substantial portion of the Galactic stellar halo has been
accreted. Today, the focus has shifted to finding out what exactly
fell onto the Milky Way and when. The reconstruction of the stellar
halo accretion history necessarily involves brushing away the Galactic
in-situ components so that the faint fragments of broken satellites
can be pieced together. Therefore, the majority of the detections of
stellar halo sub-structure have been limited to high
($|b|>30^{\circ}$) Galactic latitudes, where disk/bulge contamination
is minimal.

The current tally \citep[see e.g.][]{Be2013} indicates that as much as
$50\%-70\%$ of the stellar halo is in well-mixed, seemingly smooth
components, with another $30\%$, or perhaps even $50\%$, contributed
by four structures: the Sagittarius stream, the Galactic Anti-centre
Stellar Structure (GASS), the Virgo Cloud (containing the Virgo
Over-density and the Virgo Stellar Stream), and the Hercules-Aquila
Cloud (HAC). However, curiously two of the four most massive
structures in the stellar halo seem to reside very close to the disk
plane. The GASS covers a large region of the low-latitude sky
approximately in the direction of the anti-centre and the HAC is
hidden behind the disk and the bulge, on the other side of the Galaxy
\citep{Be2007}. Given their obvious proximity to the plane, claims
have been made as to the genesis of these structures, linking them to
perturbations in the stellar disk density. For example, \citet{La2011}
argue that while the excess of faint Main Sequence (MS) stars around
the HAC position is real, the distance to the sub-structure was
over-estimated and the true location of the Cloud is much closer, only
1 to 6 kpc from the Sun rather than 10 to 20 kpc.

\begin{figure*}
\includegraphics[width=125mm]{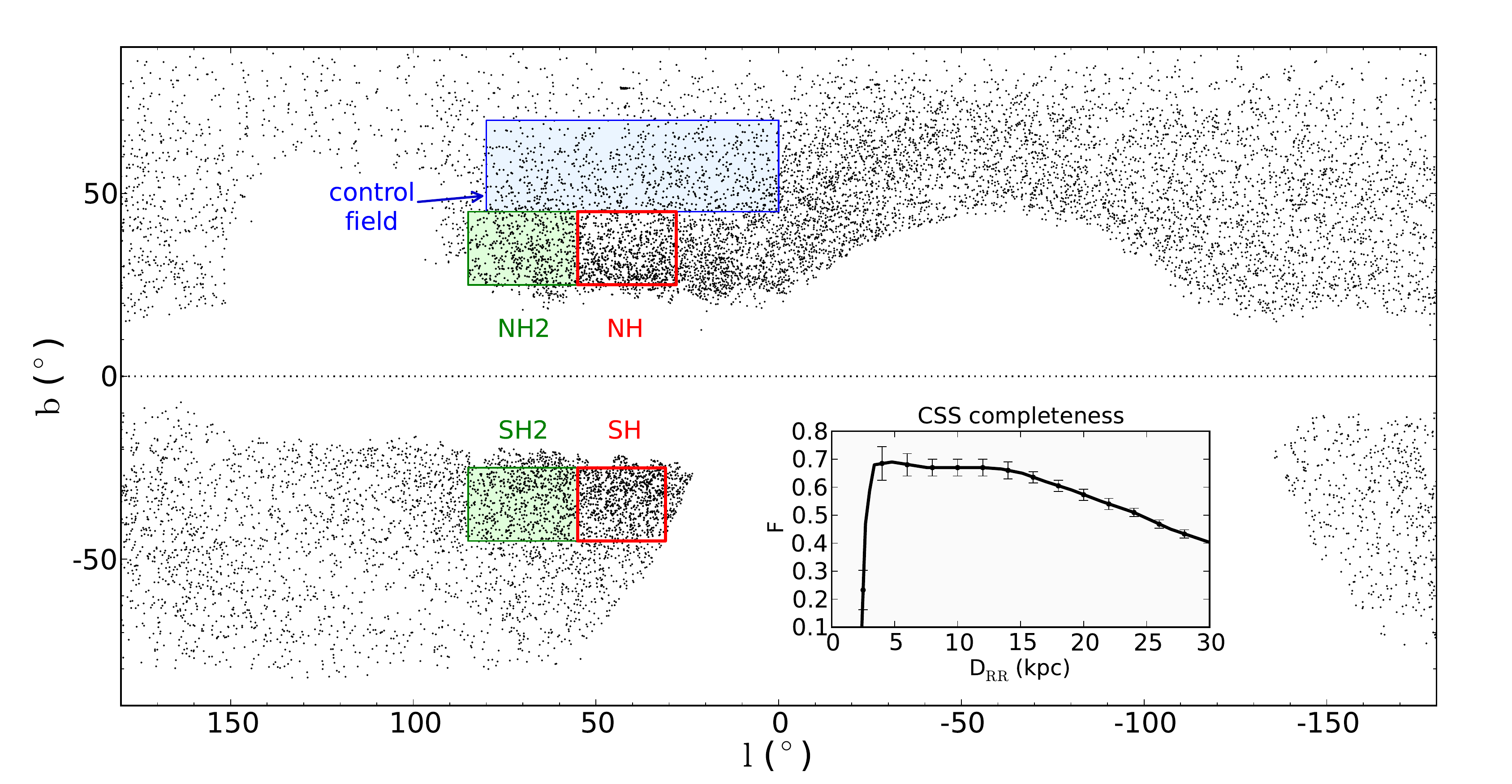}
 \caption{Spatial distribution of Catalina Schmidt Survey RRab Lyrae
   (DR13a) in Galactic coordinates. We highlight with blue our
   selection of the control field (see Section 3), with red our
   selections of the 'on-cloud' fields, in the Northern Hemisphere
   (NH) and in the Southern Hemisphere (SH) (see Section 4). In green
   we show another field selection (NH2 and SH2) that we will use in
   Section 4. The insert shows the completeness of the survey as a
   function of distance, assuming $A_{V} = 0$.}
 \label{footprints}
\end{figure*}

The HAC is a challenging structure to map. Its presence is easily
spotted due to the obvious asymmetry in the faint MS density in the
directions towards the Galactic center. However, characterizing its
full extent is far from straightforward. The Sloan Digital Sky Survey
(SDSS) data, which provides the deepest wide area view of the sky, is
not symmetric between positive and negative Galactic
longitudes and latitudes. Moreover, the contiguous coverage stops at latitudes
$|b| \approx 30^{\circ}$, but there exist some limited SDSS imaging data
along several $\sim 2.5^{\circ}$ wide stripes slicing through the
Galaxy at constant $l$ at $|b|<30^{\circ}$. The analysis of these
slices \citep{Be2007} clearly indicates that the HAC signal continues
to low latitudes, at least as low as $|b|\sim 15^{\circ}$. In
addition, the Cloud is noticeably asymmetric with respect to the disk
plane, i.e. at $l\sim50^{\circ}$, there exists a prominent
over-abundance of MS stars in the Galactic Southern hemisphere. The MS stars
contributing to the over-density are in a wide magnitude range $18 < i
< 22$, thus giving only an approximate indication of the distance to
the Cloud.

RR Lyrae are an ideal tracer population to pin down the distance and
the luminosity of the HAC. To this end, \citet{Watkins2009} and
\citet{Sesar2010} exploit the multi-epoch data in the narrow
equatorial SDSS Stripe 82 to detect RR Lyrae stars with high
efficiency. Stripe 82 briefly crosses the HAC region in the South and
both teams report a significant overdensity of RR Lyrae at
$45^{\circ}<l< 60^{\circ}$ and
$-45^{\circ}<b<-25^{\circ}$. \citet{Watkins2009} give the distance to
the HAC of 22 kpc, this is the average over the cross-section of the
Cloud and the Stripe.

In this work, we take advantage of the large sample (in excess of
10,000) of intermediate distance (5-30 kpc) RR Lyrae stars of the {\it
  ab} type detected by the Catalina Real-time Transient Survey
(CRTS). The size of the sample and the distance range probed by these
data is ample to search for the signatures of large-scale halo
sub-structure, and in particular the HAC. The paper is structured as
follows. Section 2 briefly describes the properties of the RR Lyrae
dataset analysed while Section 3 describes the smooth stellar halo
models used to reveal the Cloud and estimate its
significance. Finally, Section 4 focuses on the properties of the RR
Lyrae stars in the Cloud itself.

\section{Data}

RR Lyrae are radially pulsating horizontal branch stars, with
pulsation periods between 5 and 15 hours and amplitudes in the
range 0.2 -1.6 mag in the $V$-band. These stars have a well defined
Period-Luminosity-Metallicity relation which allows for accurate
distance determination (normally with less than 10\% uncertainty). RR
Lyrae are low-mass, long-lived stars (typically more than 10 Gyrs
old) and are found in large quantities in globular clusters and in the
stellar halo of our Galaxy making them ideal tracers of the internal
substructure of the Milky Way. They are sufficiently bright, with an
almost invariable absolute magnitude ($M_{V} \approx$ 0.59) meaning that
they can be detected out to large distances in relatively shallow
surveys, i.e. $\sim 100$ kpc for a survey with a modest limiting
magnitude between 20 and 21 mag.

RR Lyrae are also relatively abundant, with their number density in
the solar neighborhood between 4 and 6 kpc$^{-3}$. These tracers allow
for studies of the halo substructure with good spatial resolution
\citep[e.g.][]{Preston1991, Vivas2006}. Several large-scale RR Lyrae
surveys exist already and a useful comparison between them in terms of
their sky coverage and completeness is presented in Table~\ref{Table1} of
\citet{Mateu2012}.  However, 2013 saw two new RR Lyrae
catalogues published: LINEAR II \citep{Sesar2013} which probes
distances between 5 kpc and 30 kpc over $\sim$ 8,000 deg$^{2}$; and the
Catalina Schmidt Survey \citep[][hereafter DR13a,b ]{Drake2013a,
  Drake2013b} reaching distances between 2 and 60 kpc over $\sim$
20,000 deg$^{2}$ of the sky.\\
The Catalina Sky Survey began in 2004 and used three different
telescopes to discover near-Earth objects (NEOs) and potentially
hazardous asteroids (PHAs). Each of the survey telescopes is run as a
separate sub-survey. In this work, we are using the data from the
Catalina Schmidt Survey (CSS), one of the three sub-surveys of the
Catalina Sky Survey. The CSS contains an impressive sample of $\sim$
14,500 type ab RR Lyrae with magnitudes between 11.5 and 20 mag,
spread over approximately half of the sky ($0^{\circ} <$ RA $<
360^{\circ}$, $\-22^{\circ} <$ DEC $< 65^{\circ}$), the largest
Galactic volume ever surveyed with RR Lyrae. The catalogue comes with reliable distance estimates from a
well-studied absolute magnitude vs. metallicity distribution \citep{Catelan2008}.

The parameters of the CSS RRab stars are listed in Table 1 of DR13a
and Table 2 of DR13b. In what follows, we are going to use their
published equatorial coordinates RA, DEC, (see
Figure~\ref{footprints}), as well as the magnitudes $V_{0}$ and the
heliocentric distances $D_{RR}$. The $V_{0}$ magnitudes are already
extinction corrected using the \citet{Schlegel1998} reddening maps.
The uncertainties on all of the above quantities are not listed but
estimates are provided. The average metallicity of the sub-sample with
available SDSS spectra is approximately $-1.55$ (see Figure 20, DR13a)
therefore, given that the absolute magnitude of the RRL has a linear
dependence on metallicity \citep[e.g.][]{Chaboyer1999}, we can assume
$M_{V} \approx 0.6 $ for the CSS data. The dispersion in the
metallicity of the sub-sample, $\sim$ 0.3 dex, introduces an error on
the absolute magnitude, while the apparent magnitudes are affected by
photometric error. These two factors translate into a total
uncertainty on the distances of the order of 7\% (e.g. at $D = 15$
kpc the error will be $\delta D \approx \pm$ 1 kpc).
 \setcounter{table}{0}
\begin{table*}
 \centering
 \begin{minipage}{140mm}
  \caption{Single power-law and broken power-laws models with their
    parameters. The normalisation for the \citet{Watkins2009} model is
    divided by 11, in agreement with \citet{Sesar2010} and DR13b.}
 \label{Table1}
  \begin{tabular}{@{}lccclccccr@{}}
 \hline
   Model & $\alpha_{in}$ & $\alpha_{out}$ & $ r_{b}$&$ q_{H}$ & $\rho_{\odot}^{RR}$ & No of tracers & Area & Survey &Reference \\
              & & & & & kpc$^{-3}$ & (approx) & deg$^{2}$ &  & \\
 \hline
   BPL & 2.40 & 4.50 & 23 &1.00 & 3.0 & 316 RRab & 290 & Stripe 82 & \citet{Watkins2009} \\
     SPL & 2.77 &  &  & 0.64 & 5.6& 366 RRab & 290 & Stripe 82 & \citet{Sesar2010} \\
  BPL& 2.30 & 4.60& 27 &0.61 & \textit{7.3}&  6,800 BHBs & 14,000 & SDSS DR8 & \citet{Deason2011} \\
  BPL  & 2.62 & 3.80 & 28 &0.71 & 5.9 &  34,000 MS & 170 & CFHTLS & \citet{Sesar2011} \\
   SPL & 2.42 & & & 0.63 & 5.6 &  4,000 RRab & 8,000 & LINEAR II & \citet{Sesar2013} \\
  \hline
 \end{tabular}
 \end{minipage}
\end{table*}
\begin{figure*}
\includegraphics[width=82mm]{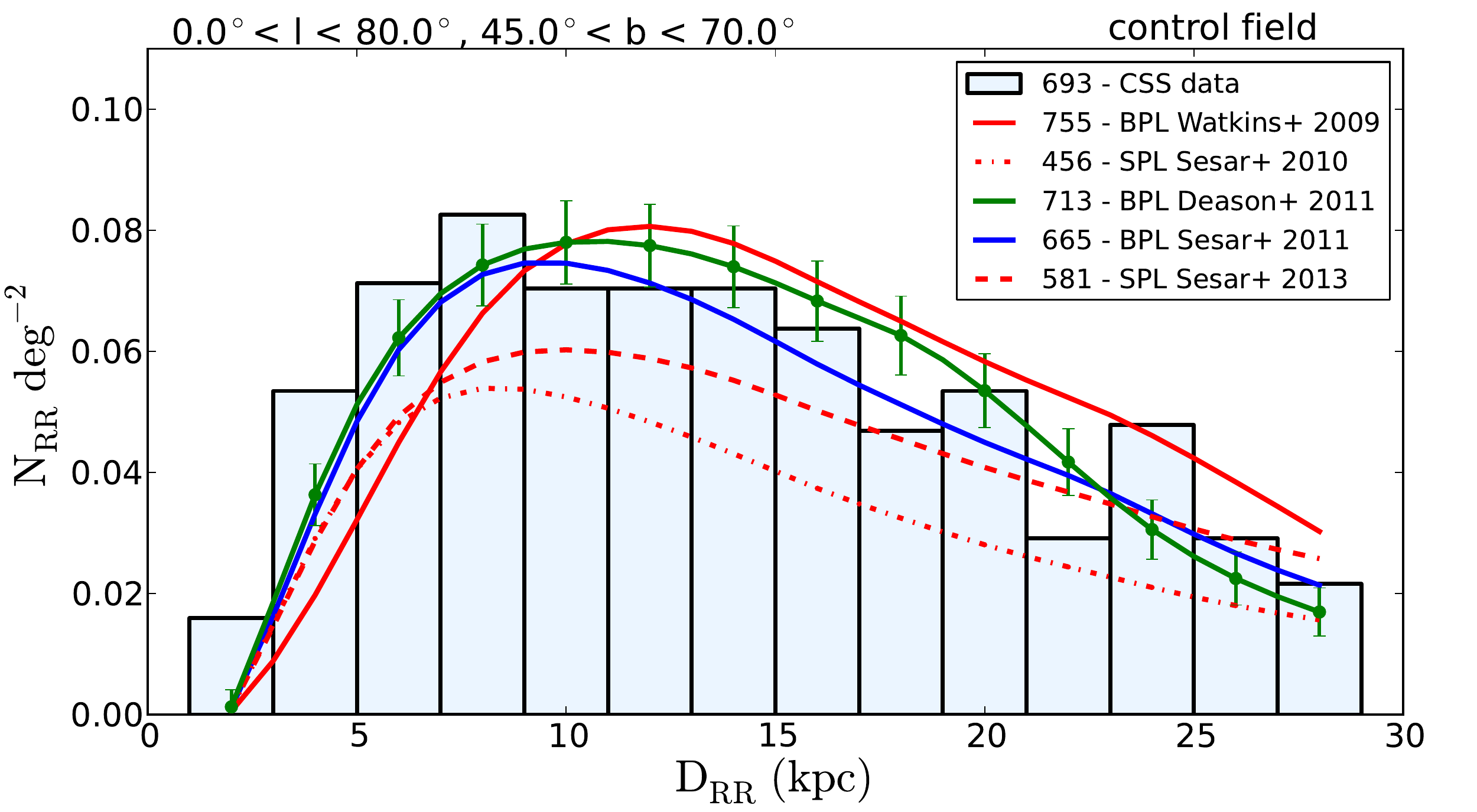}
\includegraphics[width=82mm]{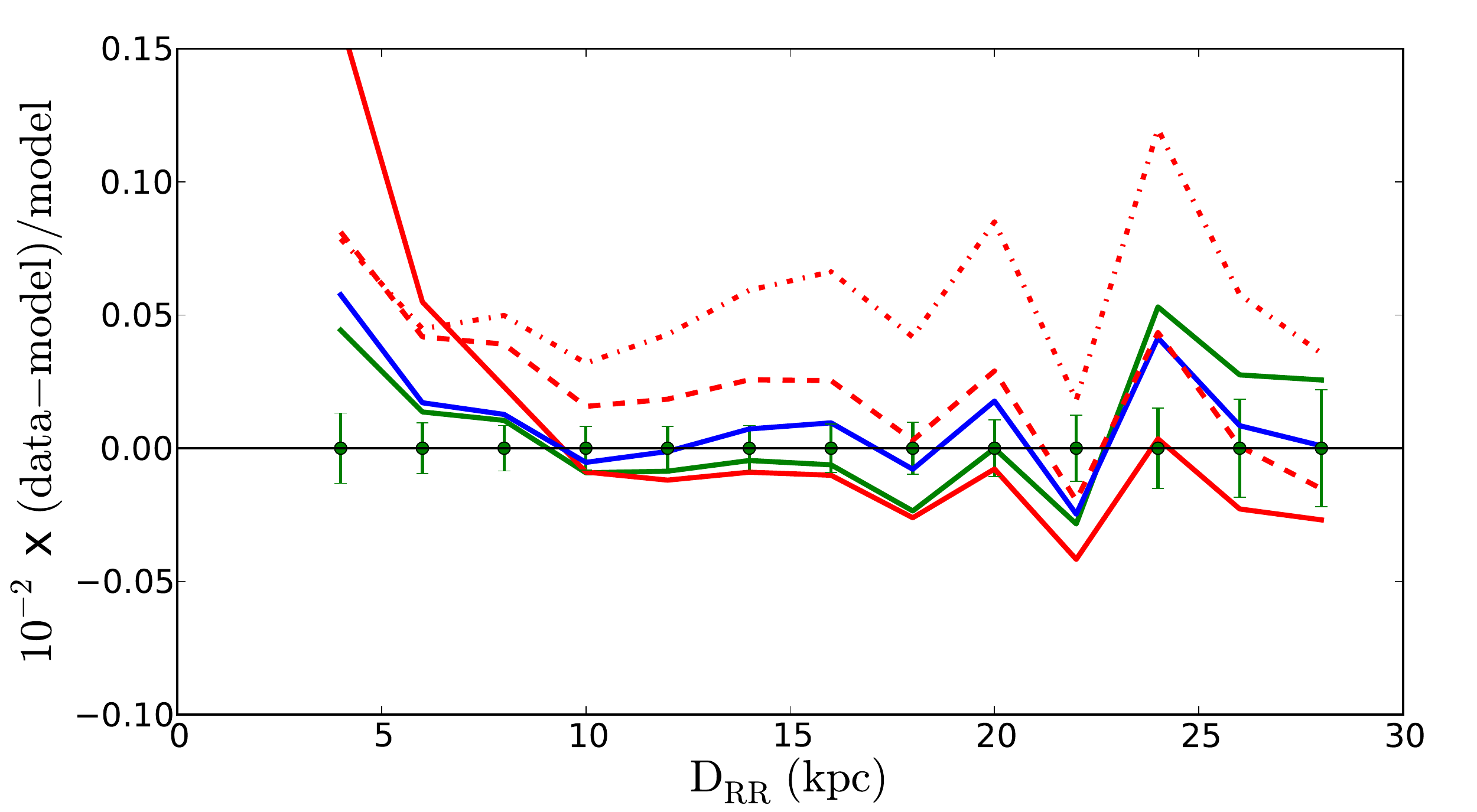}
\caption{Left panel: Comparison between the heliocentric distance number density distribution of CSS RRab Lyrae in the control field (selected in
  blue in Figure~\ref{footprints}) and five simple power law model predictions (see labels and Table~\ref{Table1}). The numbers in the legend are the
  integrated number of RRab in the 5-28 kpc distance range. As our
  sample of RRab is incomplete, we corrected the model predictions for
  the effects of completeness, using equation~\ref{compl}. Right panel: Relative difference between the data and the model. The most suitable models for predicting the data in the control field are the \citet{Deason2011} and \citet{Sesar2011} broken power-law models (the \citet{Deason2011} model was scaled such that it reproduces the number of RRL observed in this field).}
 \label{control-field}
\end{figure*} 
%
\section{Simple smooth halo models}

To probe the clumpiness of the Galactic RR Lyrae distribution on the
scales of several kpc, a reasonable first step is to compare the
observed tracer volume density to the predictions of a simple smooth
halo model. To guide this comparison we identify several locations of
interest in the central Galaxy.  Figure~\ref{footprints} shows the
spatial distribution of all $\sim$14,400 RRab Lyrae in the CSS
sample. The principal ``on-cloud'' fields in the South and North
Galactic hemispheres are plotted in red and marked with NH and SH
respectively, the secondary fields are marked in green. Additionally,
a comparison field (control field) is chosen at suitably high Galactic
latitude in the region free of known sub-structure; this appears in
blue. These fields are selected with the following logic in mind. In
the comparison field, a smooth density model can be found that
describes the data well. In the primary ``on-cloud'' fields, the
strength of the residuals with respect to the model can be used to
ascertain the presence of the Cloud. The neighboring secondary fields,
labelled NH2 and SH2, are analysed to constrain the extent of the HAC.

As illustrated in e.g. \citet{Deason2011}, power law volume density
models are fairly successful at predicting the number counts of old
metal-poor stars in the Galactic halo. According to these authors,
within 20 kpc of the Galactic centre stellar halo substructure, as
traced by BHBs, reaches the levels of $10\%-20\%$ when the known large
overdensities such as the Sagittarius stream are removed. Since, apart
from the HAC, there are no other halo structures known in the region
we are focusing on, it is safe to posit that similar levels of
smoothness should be seen in the RRab tracers, at least in the
comparison field. This is not an unreasonable assumption to make given
the earlier successful RR Lyrae modelling attempts, though with
admittedly smaller datasets \citep[e.g.][]{Watkins2009, Sesar2010,
  Sesar2011}.

Simple stellar number count models of the halo are based on the following
formalism.
The heliocentric distance $D$ and galactic coordinates $(l,b)$ of a
star and its Galactocentric Cartesian coordinates are related as
\begin{equation}
\begin{aligned}
X= D \cos(l)\cos(b) - R_{\odot} \\ \nonumber 
Y=D\sin(l)\cos(b) \nonumber  \\
Z=D\sin(b)  
\\
r^{2} = X^{2} + Y^{2} + Z^{2} q_{H}^{-2} 
\end{aligned}
\end{equation}
where $q_H$ describes the flattening of the halo.

\noindent Here $R_{\odot}$ is the Sun's distance from the Galactic
center and $r$ is the star's Galactocentric ellipsoidal distance. The
Sun is at $(X, Y, Z)$ = (-8,0,0) kpc, the X-axis points towards the
Galactic center, Y component in the direction of Galactic rotation and
the Z-axis points toward the north Galactic pole.

The stellar density distribution in the inner halo (within $\sim 20$
kpc) appears to be well described with a simple power-law form
\begin{equation}
\rho_{model}^{RR}(D,l,b) =
\rho_{\odot}^{RR}\bigg(\frac{R_{\odot}}{r}\bigg)^{\alpha}.
\end{equation}

\noindent At larger distances, a steeper power-law index is required to
fit the data well \citep[see e.g.][]{Deason2011}, therefore leading to
an overall broken power-law model with characteristic ``break radius''
$r_{b}$

\begin{figure*}
\hspace*{-0.05cm}
\includegraphics[width=58.1mm, height = 35mm]{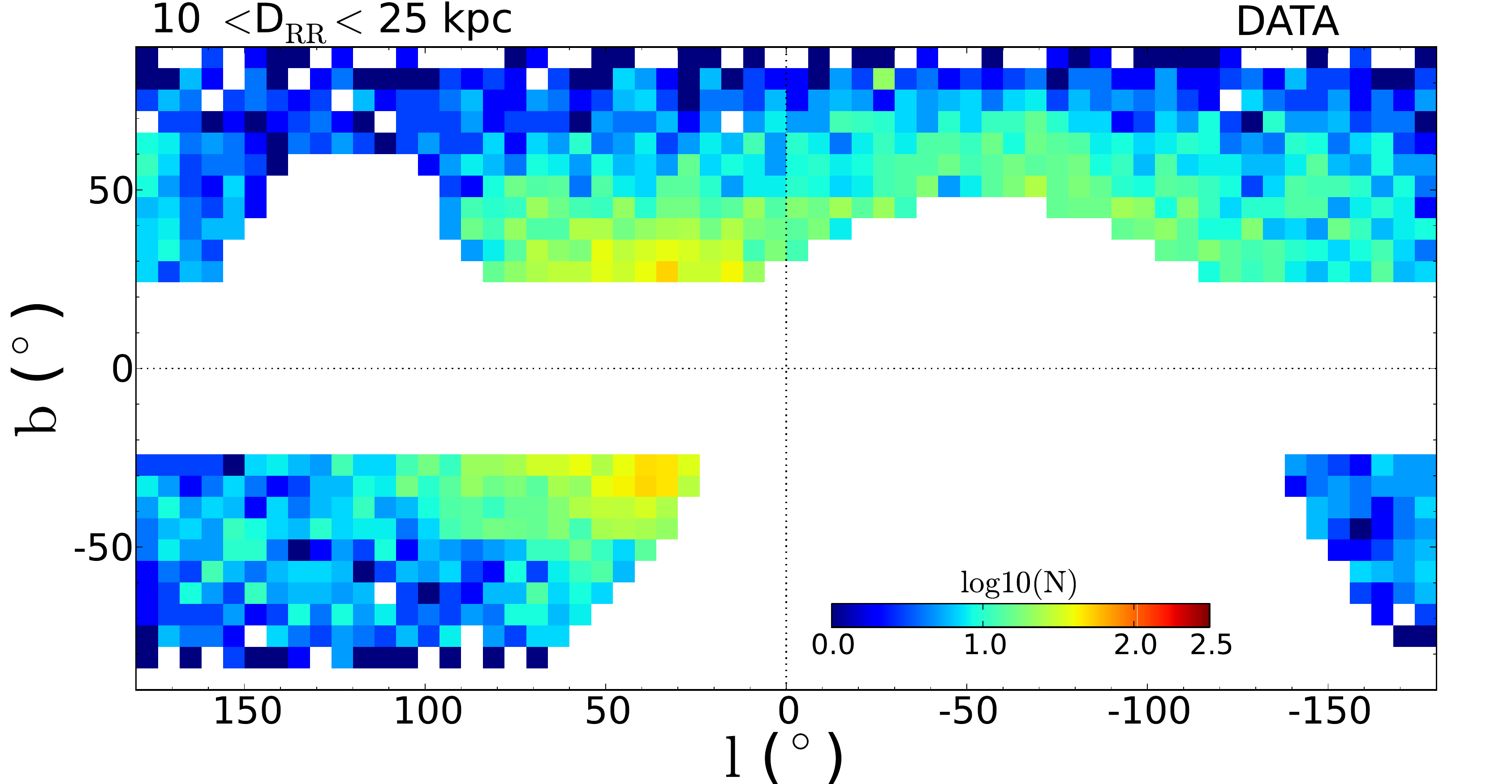}
\includegraphics[width=58.1mm, height = 35mm]{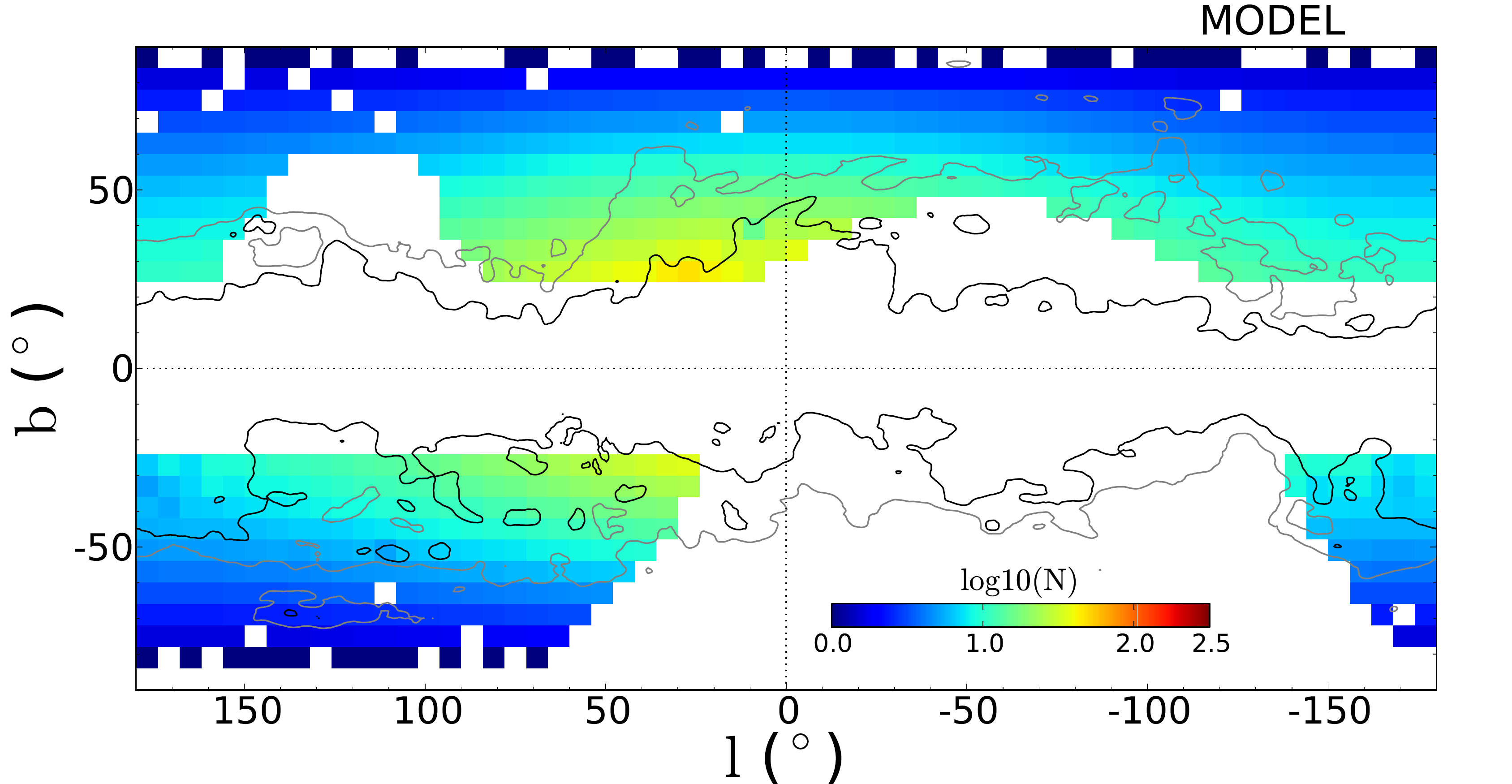}
\includegraphics[width=58.1mm,height = 35mm]{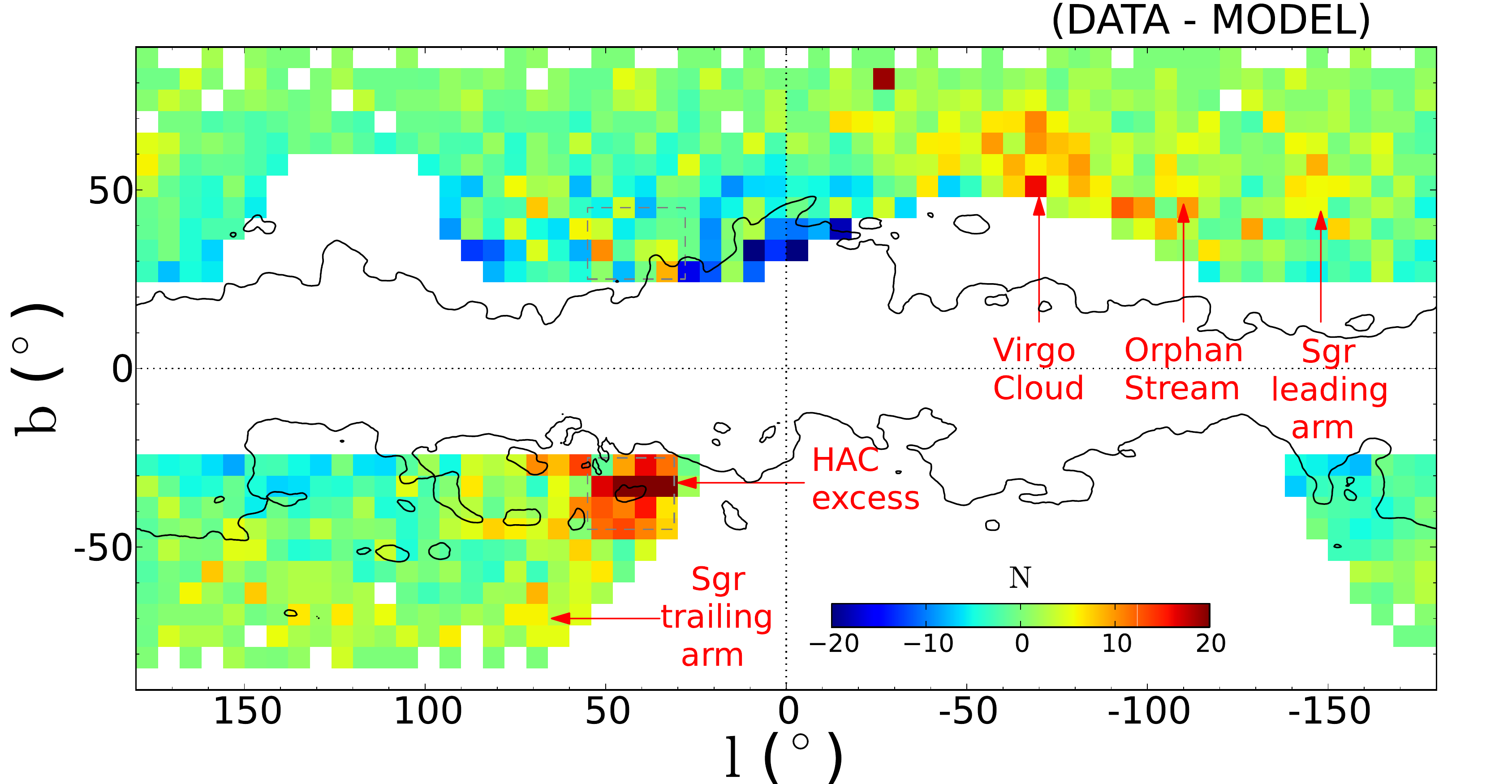}\\
\includegraphics[width=58mm, height = 36mm]{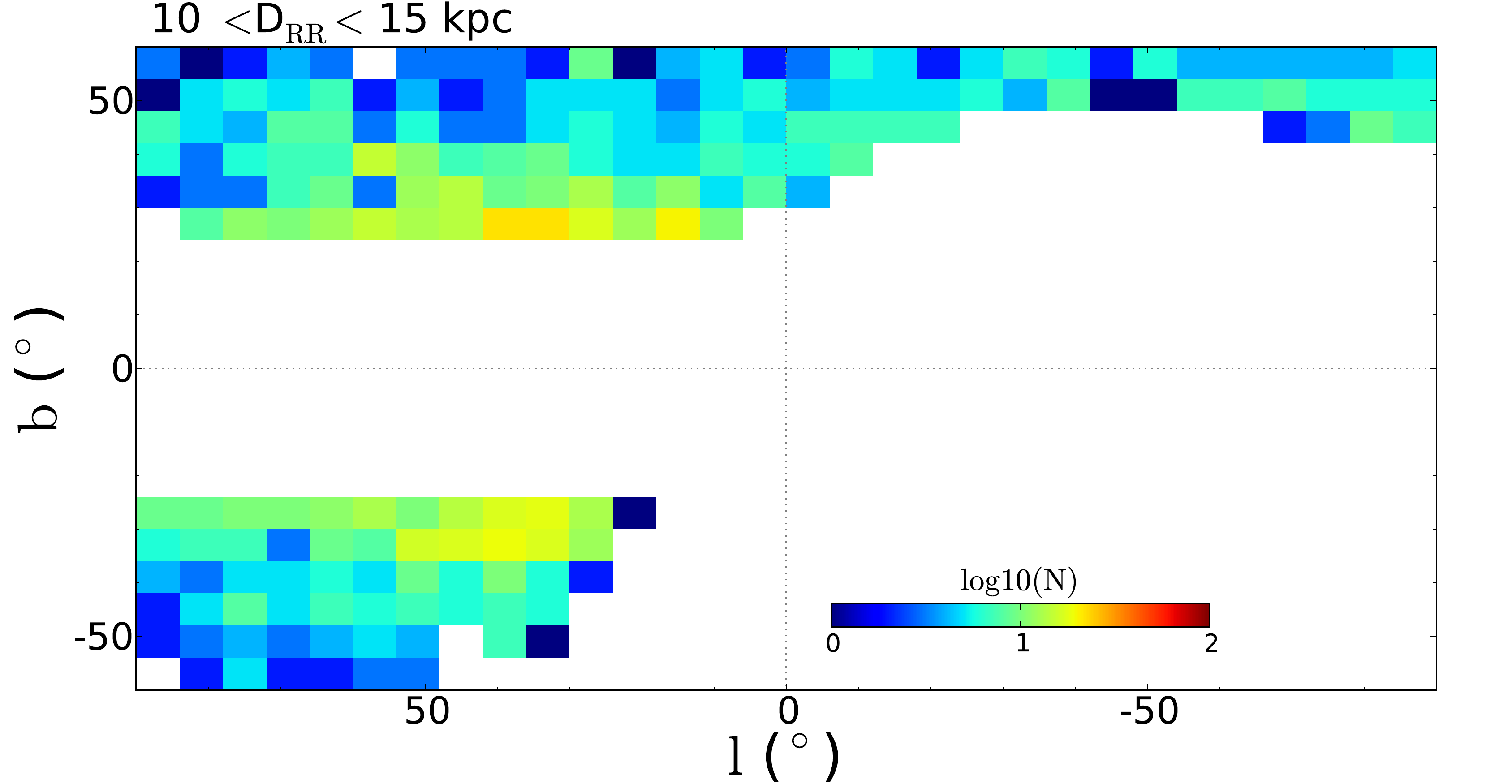}
\includegraphics[width=58mm, height = 36mm]{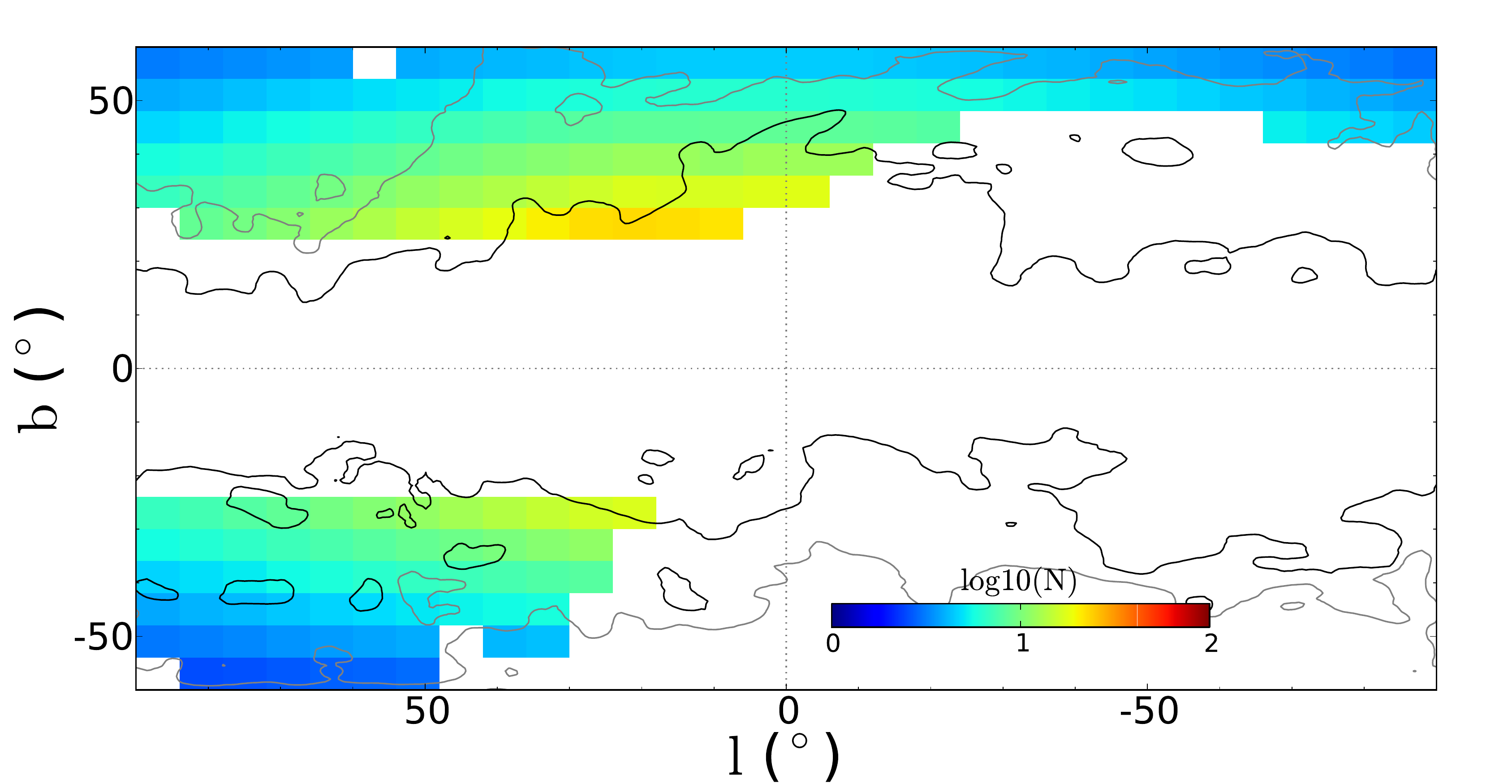}
\includegraphics[width=58mm, height = 36mm]{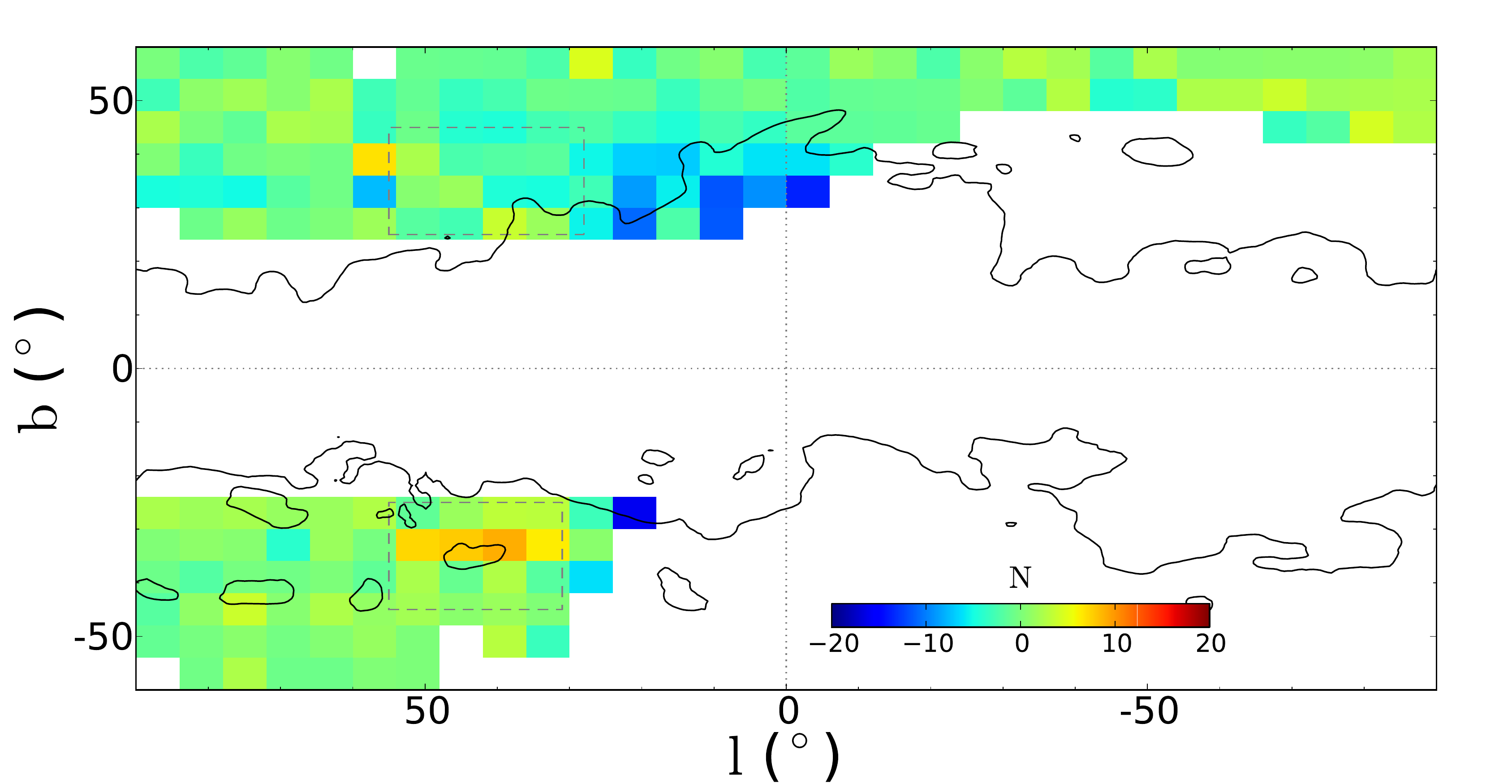}
\includegraphics[width=58mm, height = 36mm]{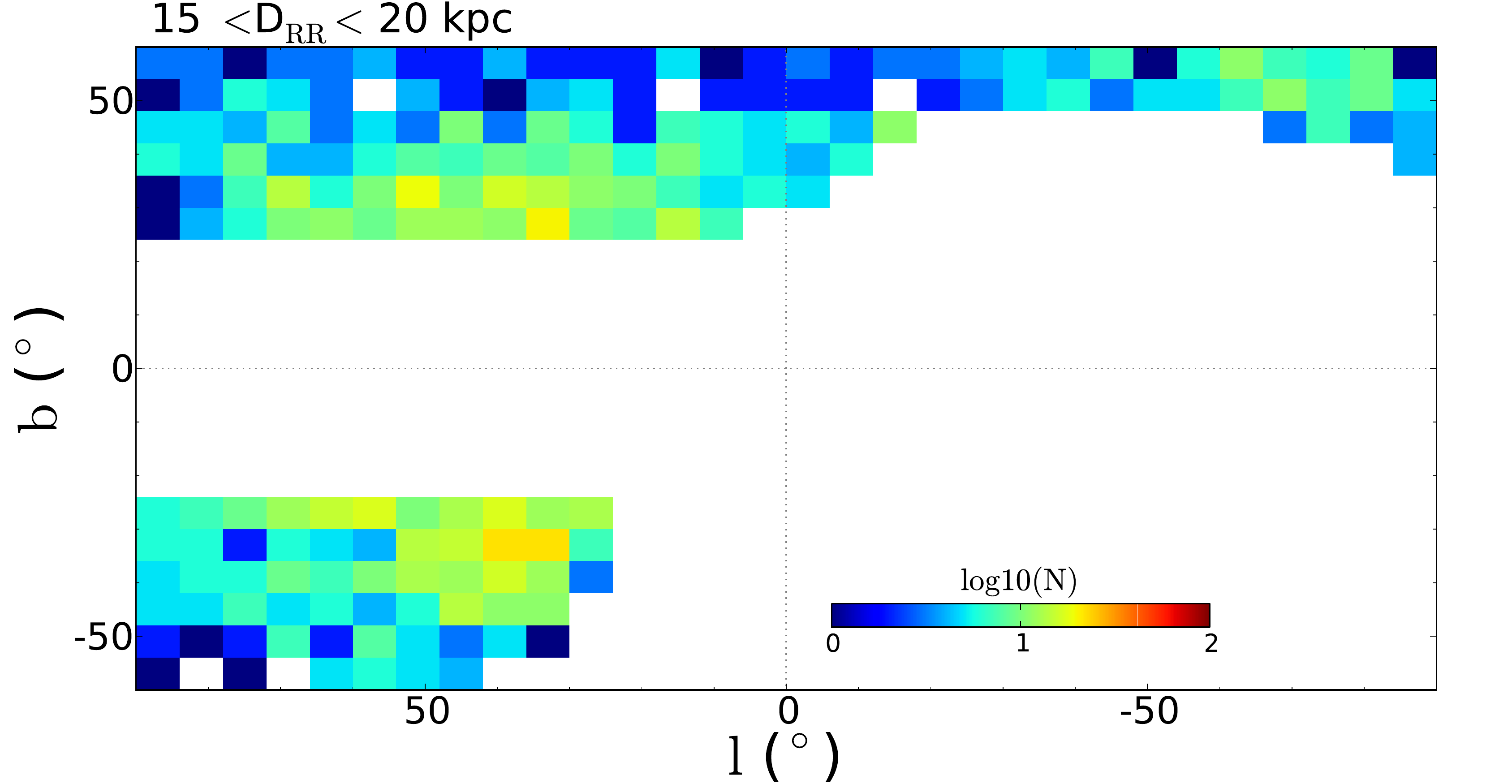}
\includegraphics[width=58mm, height = 36mm]{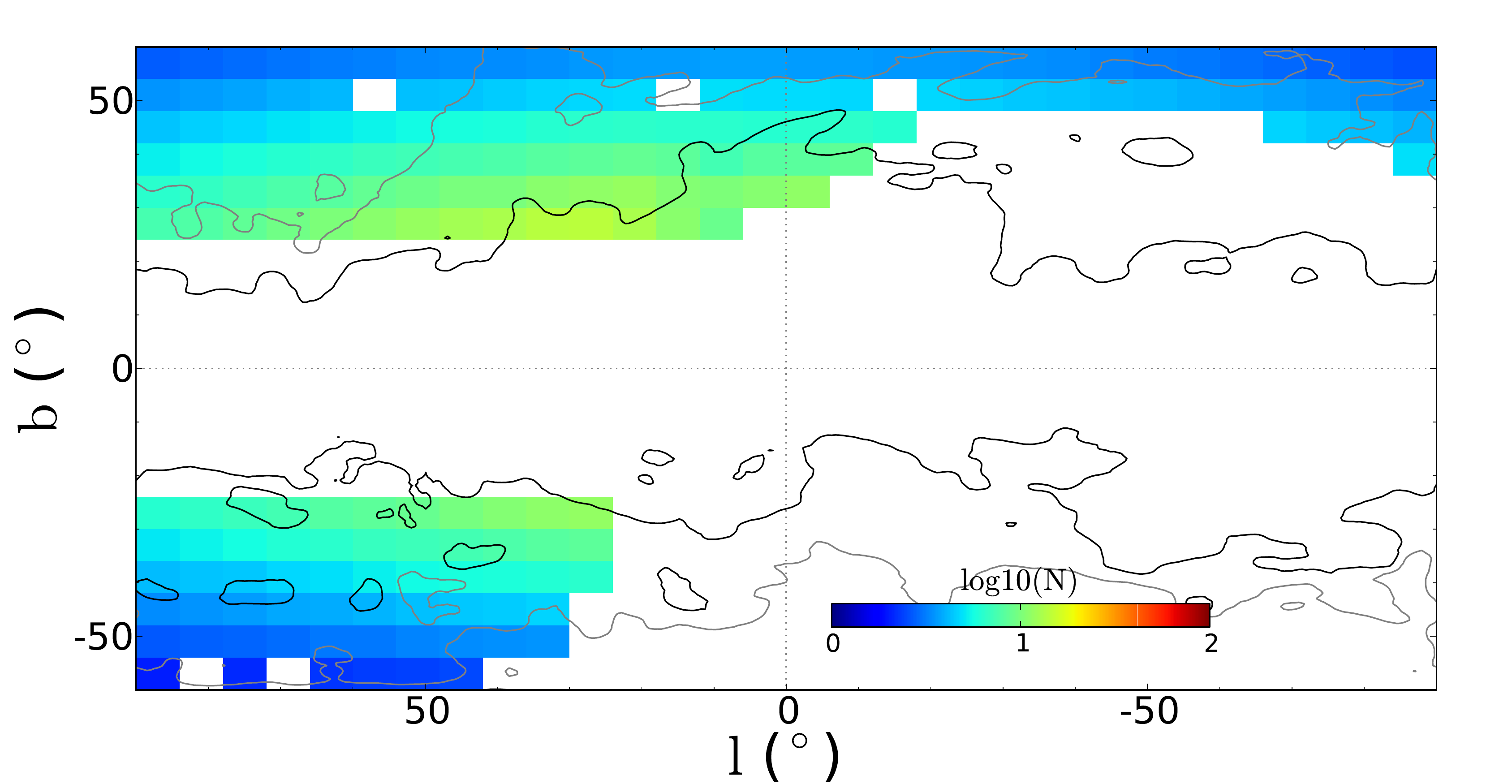}
\includegraphics[width=58mm, height = 36mm]{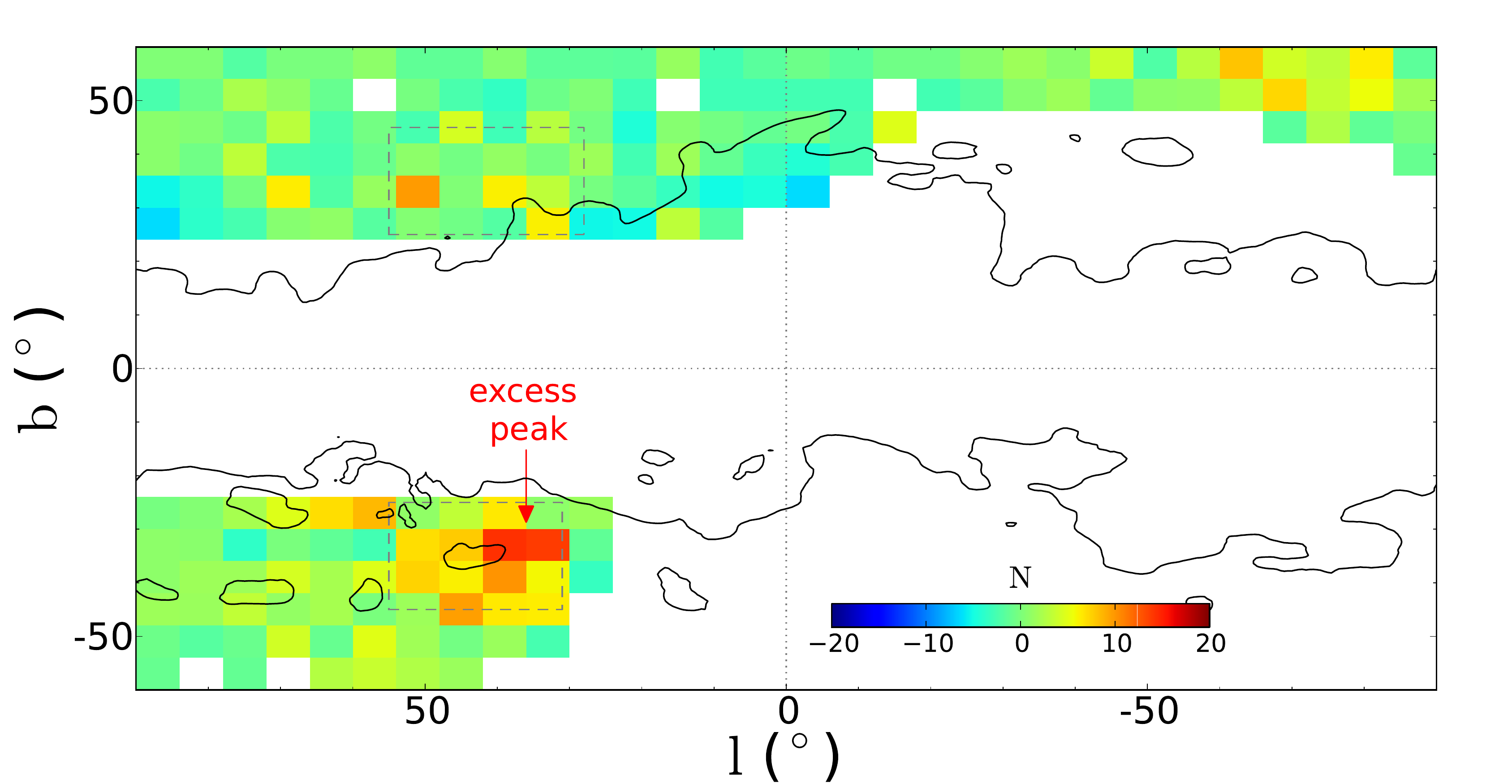}
\includegraphics[width=58mm, height = 36mm]{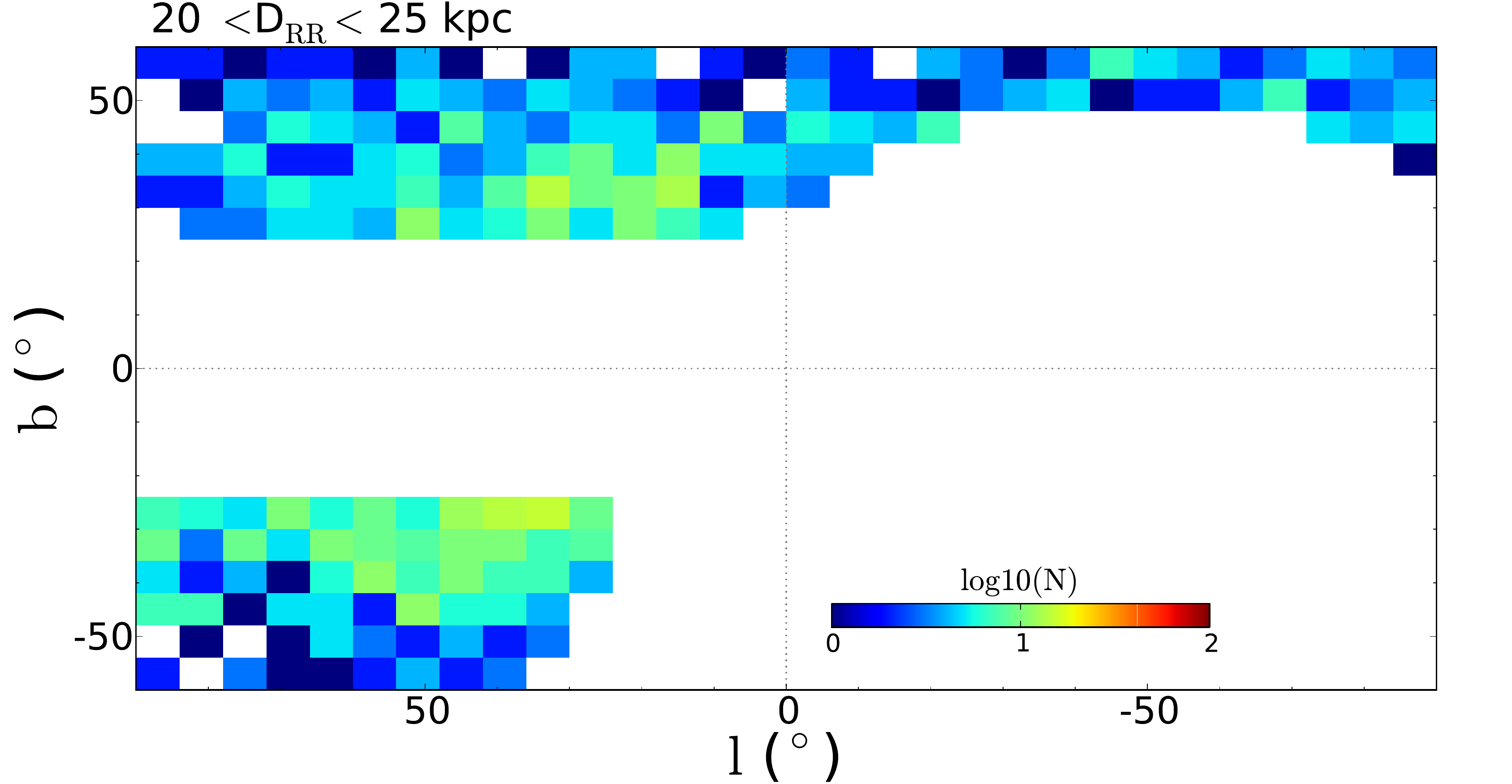}
\includegraphics[width=58mm, height = 36mm]{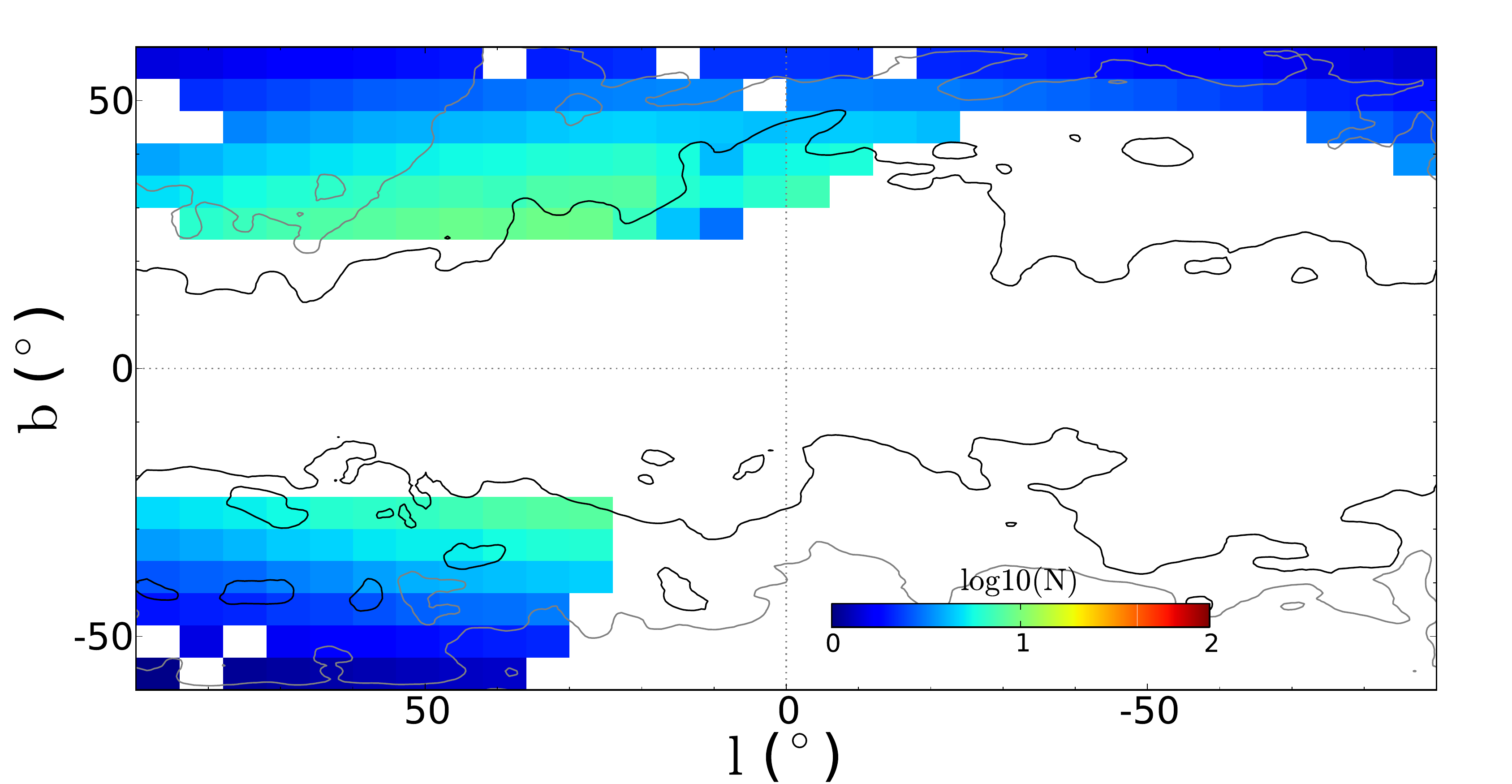}
\includegraphics[width=58mm, height = 38mm]{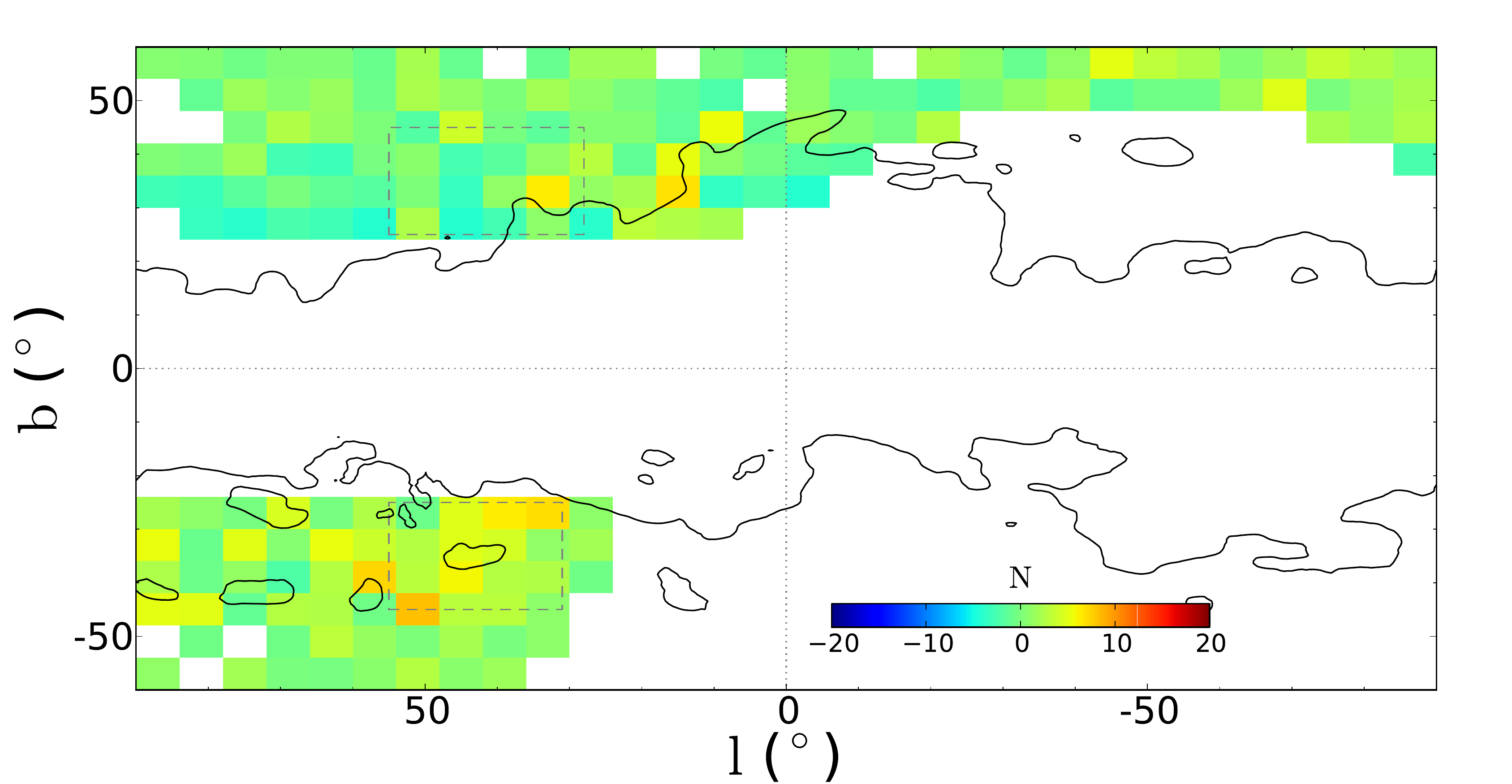}
\caption{Data, model predictions and residuals (left, middle and right
  panels respectively) in Galactic coordinates, in $6^{\circ}$ x
  $6^{\circ}$ bins. The number density map was not smoothed and it
  takes into account the efficiency correction at varying $l$ and
  $b$. Overlaid on the plots are contours of constant reddening E(B-V)
  (black is E(B-V) = 0.10 and gray is E(B-V) = 0.04).  Top panels:
  all-sky density distribution of RR Lyrae with $10 < D_{RR} < 25$ kpc
  in the CSS (left) and the model \citet{Deason2011} (middle). The
  residuals (right) highlight the halo substructure, in particular the
  Virgo Cloud (possibly mostly the VSS), the Orphan Stream, and the
  Sagittarius stream. A strong excess of RR Lyrae is present in the SH
  region. Lower panels: zoom-in density maps of the central
  $180^{\circ}$ of the Galaxy in three distance bins: $10 < D_{RR} <
  15$ kpc (2nd row), $15 < D_{RR} < 20$ kpc (3rd row) and $20 < D_{RR}
  < 25$ kpc (4th row). The peak of the excess in the SH field is
  located in the $15 < D_{RR} < 20$ kpc distance bin, as also shown in
  Figure~\ref{RRL-HAC}. Note that the left and middle panels show the
  logarithm of the stellar number density in each $6$ x $6$ deg$^{2}$
  bin.}

\label{lbmap}
\end{figure*} 
\begin{equation}
\delimiterfactor=1200 
\rho_{model}^{RR}(r) = \rho_{\odot}^{RR} (\frac{R_{\odot}}{r_{b}})^{\alpha_{in}} \times \left\{%
\begin{array}{ll}
(\frac{r_{b}}{r})^{{\alpha}_{in}}  &\textrm{if }r \leq r_{b},\\
\\
(\frac{r_{b}}{r})^{{\alpha}_{out}}  &\textrm{if }r > r_{b}.
\end{array}%
\right.
\end{equation}

\noindent For a given density model, we can estimate the number of RR
Lyrae in the given solid angle within an increment of heliocentric
distance $\Delta D_{RR}$ around $D_{RR}$
\begin{equation}
\begin{aligned}
\Delta N_{exp}(D_{RR},l,b) = \rho_{\odot}^{RR} D_{RR}^{2}
\rho_{model}^{RR}(D_{RR}, l, b) \\ \cos b \Delta D_{RR} \Delta l \Delta
b,
\end{aligned}
\end{equation}
where $\rho_{\odot}^{RR}$ is number density of RRab in the solar
neighborhood.

\begin{figure*}
\vspace{-0.4cm}
\includegraphics[width=75mm]{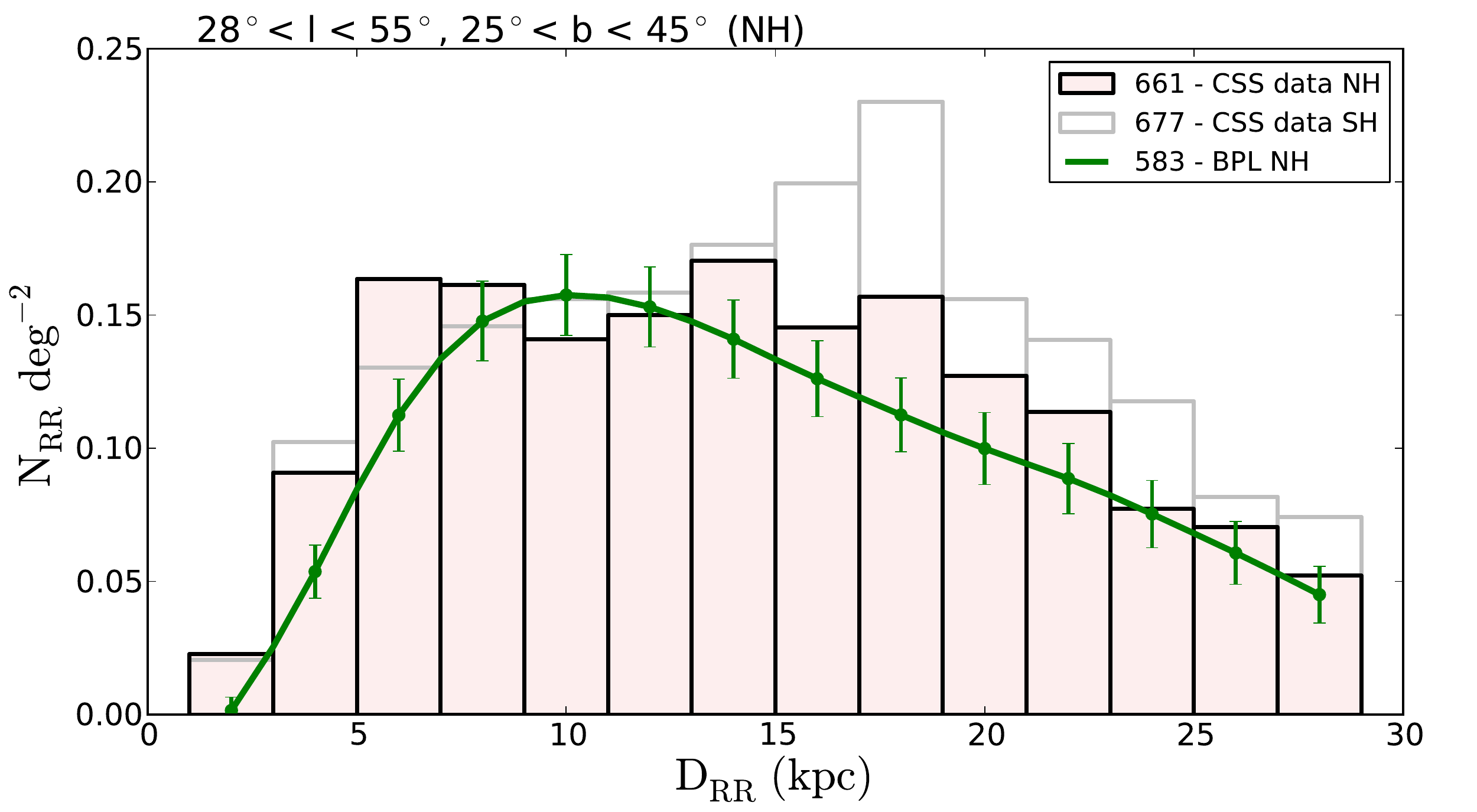}
\includegraphics[width=75mm]{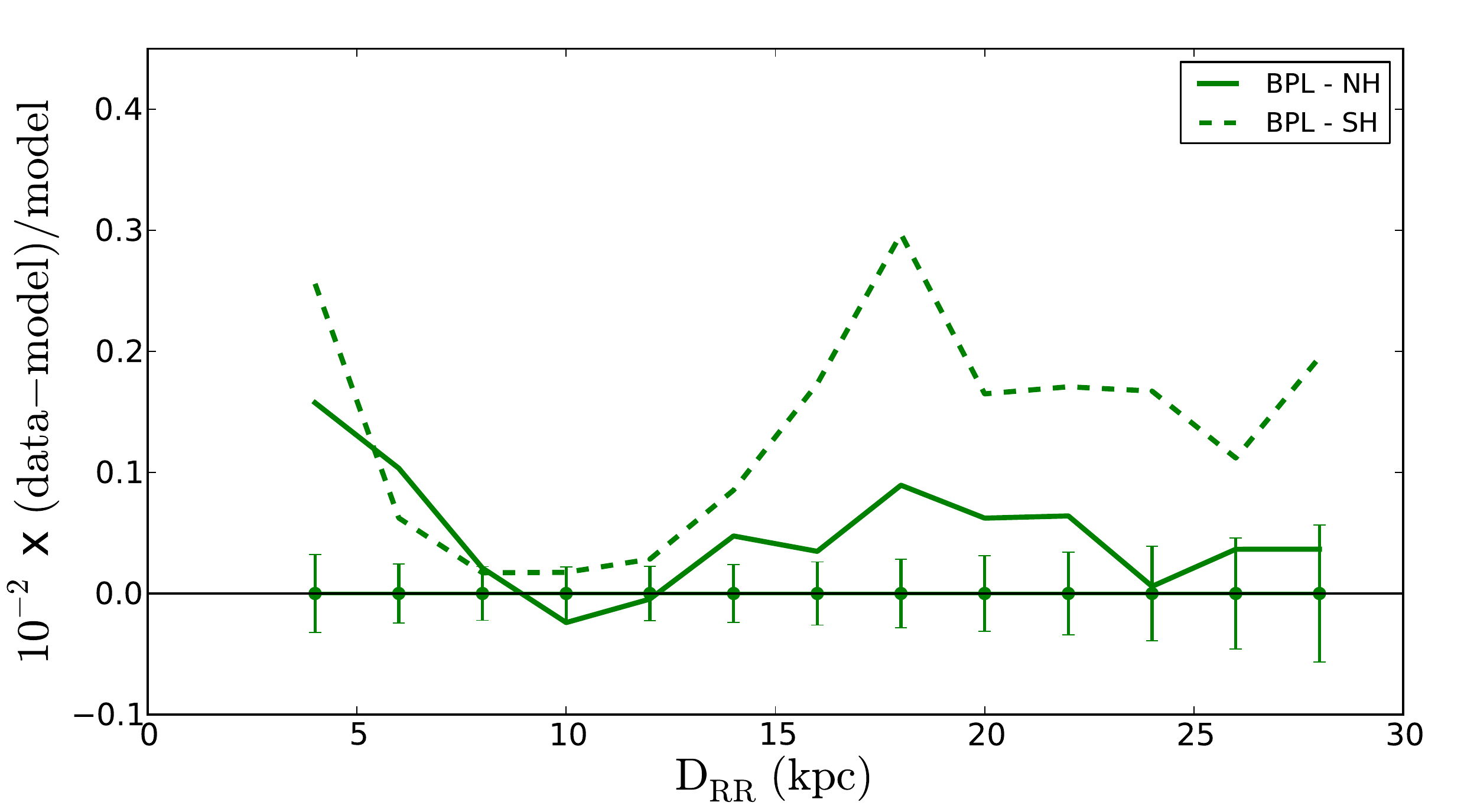}
\includegraphics[width=75mm]{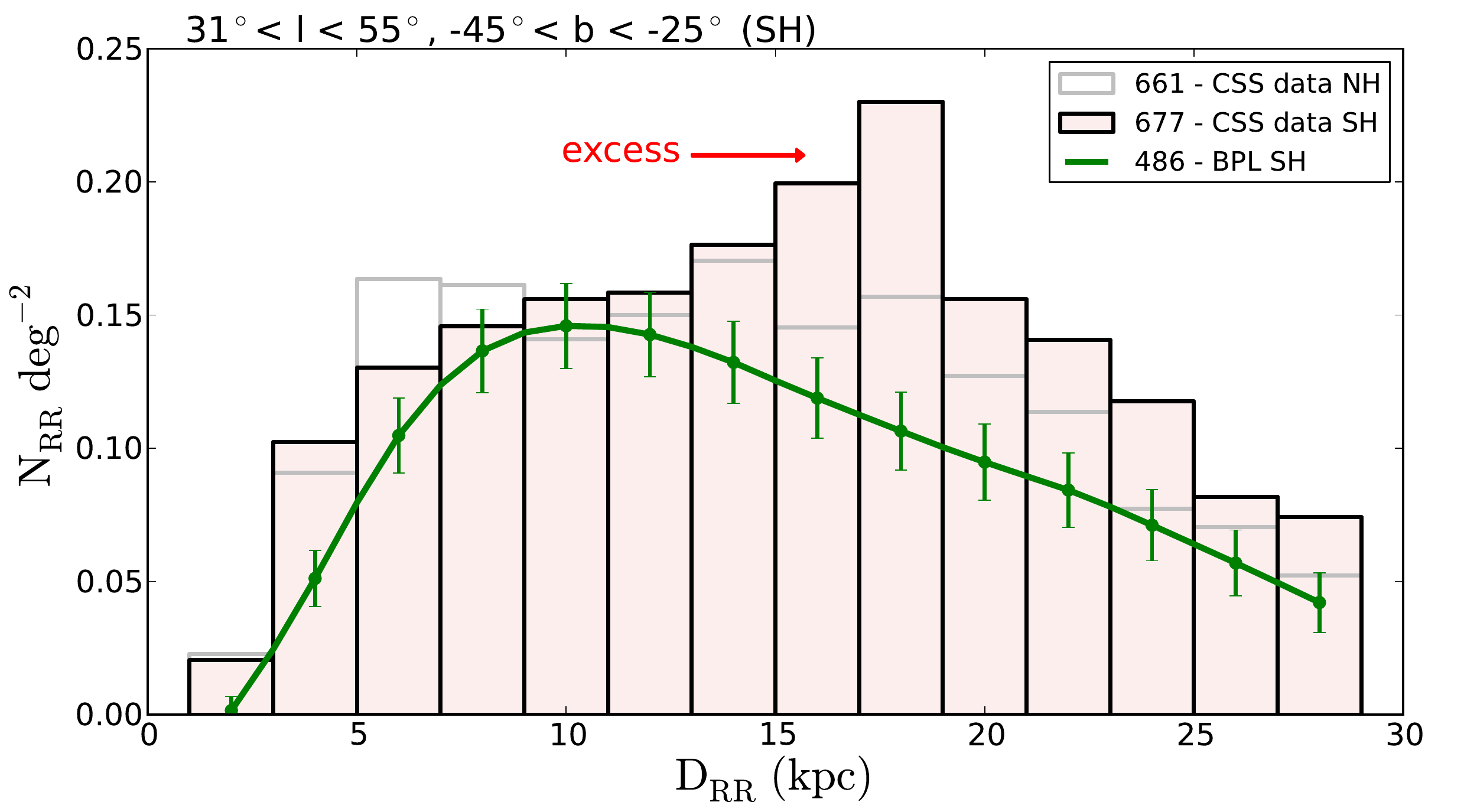}
\includegraphics[width=75mm]{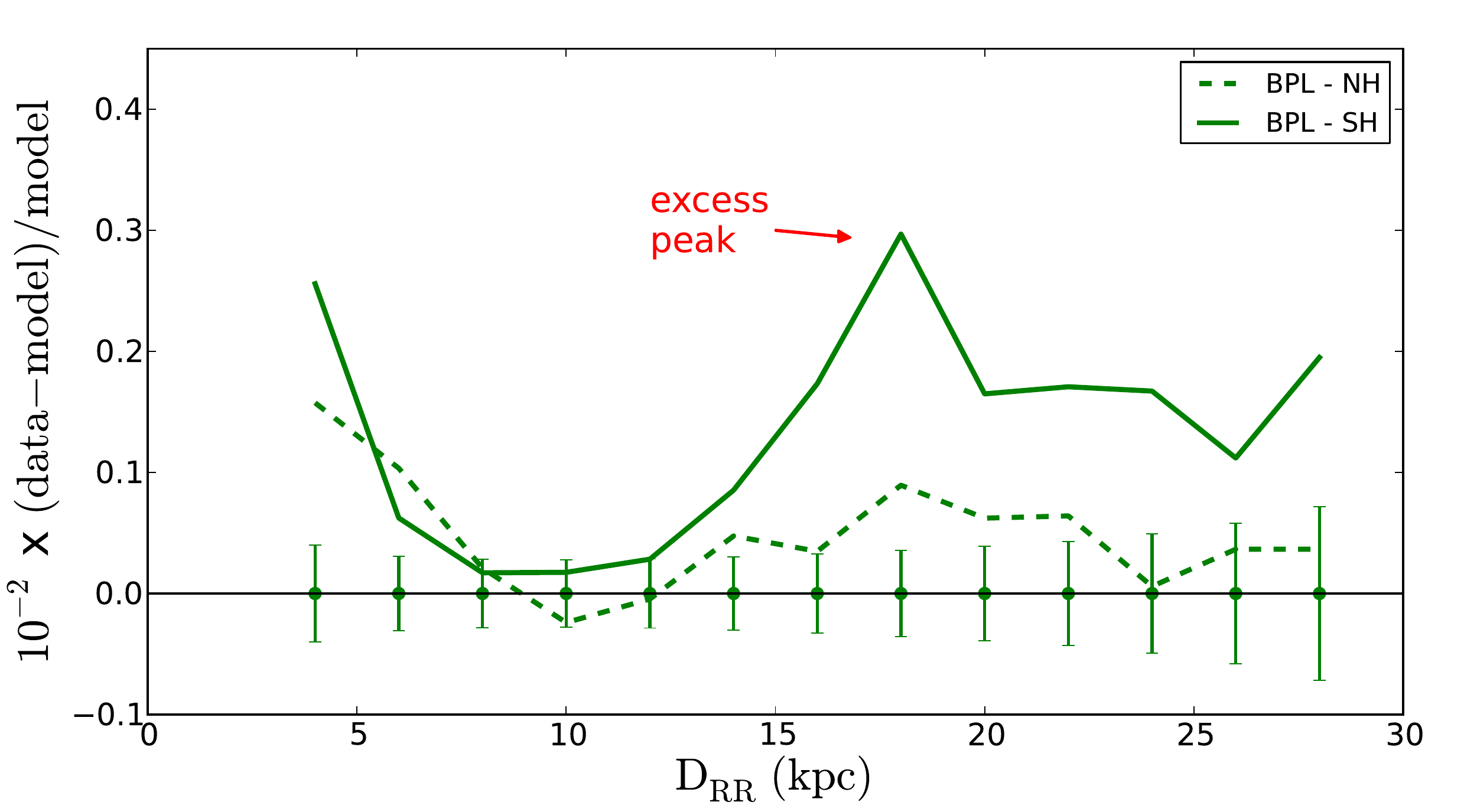}
\caption{Left panels: heliocentric distance number density
  distribution for CSS RRab in bins of 2 kpc and \citet{Deason2011}
  model predictions with error bars (in green), completeness-corrected, 
  in the NH and SH fields (red selection in
  Figure~\ref{footprints}). Right panels: Residuals between the data
  and the model. The star density distributions are asymmetric in the
  NH and SH fields, with a strong excess of stars in the SH between 12
  and 25 kpc and a peak at $\sim$ 18 kpc, also revealed in
  Figure~\ref{lbmap}. The excess in the NH field is weaker and not
  particularly significant taking into account the large error
  bars (see legend for plot labels).}

 \label{RRL-HAC}
\end{figure*} 

As the CSS RRab catalogue is incomplete, with the completeness
dropping at faint magnitudes from under $70\%$ to under $40\%$, we
need to scale down the model predictions (and more so at large
distances) to take the survey detection efficiency into account. The
fraction of RR Lyrae detected by the CSS as a function of magnitude is
presented in Figure 13 of DR13a. The insert of Figure~\ref{footprints}
shows the completeness as a function of heliocentric distance. This
curve is derived from Figure 13 of DR13a, using the well-known
relation between apparent magnitude and distance modulus and assuming
$M_{V}= 0.6$:
\begin{equation}
D_{RR} = 10^{0.2(V - M_{V} + 5 - A_{V})}
\end{equation}
where $V_{0} = V - A_{V}$ is the average magnitude from the Fourier fit to the
RRab light curves listed in Table 1 of DR13a and Table 2 of DR13b. The
efficiency-corrected number of stars predicted by the model is thus
given by
\begin{equation}
\Delta N_{obs}(D_{RR},l,b)  = F (D_{RR}) \Delta N_{exp} (D_{RR},l,b) 
\label{compl}
\end{equation}
where $F$ is the completeness function shown in
Figure~\ref{footprints}. The efficiency curve needs to be shifted in distance as $A_{V}$ varies.

Rather than fitting our own density model $\rho_{model}$ to
the CSS RR Lyrae data we have re-used some of the halo model parameters 
obtained in earlier studies.\footnote{For a discussion
of the difficulties of fitting a smooth density distribution to an RR
Lyrae dataset, see section 7 of \citet{Sesar2013}.} 
Table 1 lists the best-fit values for the number density
distribution of different halo tracers: main-sequence stars
\citep{Sesar2010}; RRab Lyrae between 5 and 23 kpc \citep{Sesar2013};
and blue horizontal branch stars \citep{Deason2011}.  Also displayed
are the values obtained in two independent studies of the RR Lyrae
density distribution in Stripe 82 \citep{Watkins2009, Sesar2010}. All
normalisation values $\rho_{\odot}^{RR}$ listed in Table~\ref{Table1}
are estimated for the Milky Way halo RRab stars. Some of these were
obtained using the actual RR Lyrae tracers, for example those by
\citet{Watkins2009} and \citet{Sesar2010} detected in the SDSS Stripe
82 data and the LINEAR II catalogue \citep{Sesar2013}. However, in two
out of five modeling efforts, stars other than RR Lyrae were employed.
For the BPL model fit to MS stars from the Canada
France Hawaii Telescope Legacy Survey, Table~\ref{Table1} gives the
conversion of the density normalisation to RR Lyrae counts derived by
\citet{Sesar2013}.

\begin{figure*}
 \includegraphics[width=155mm]{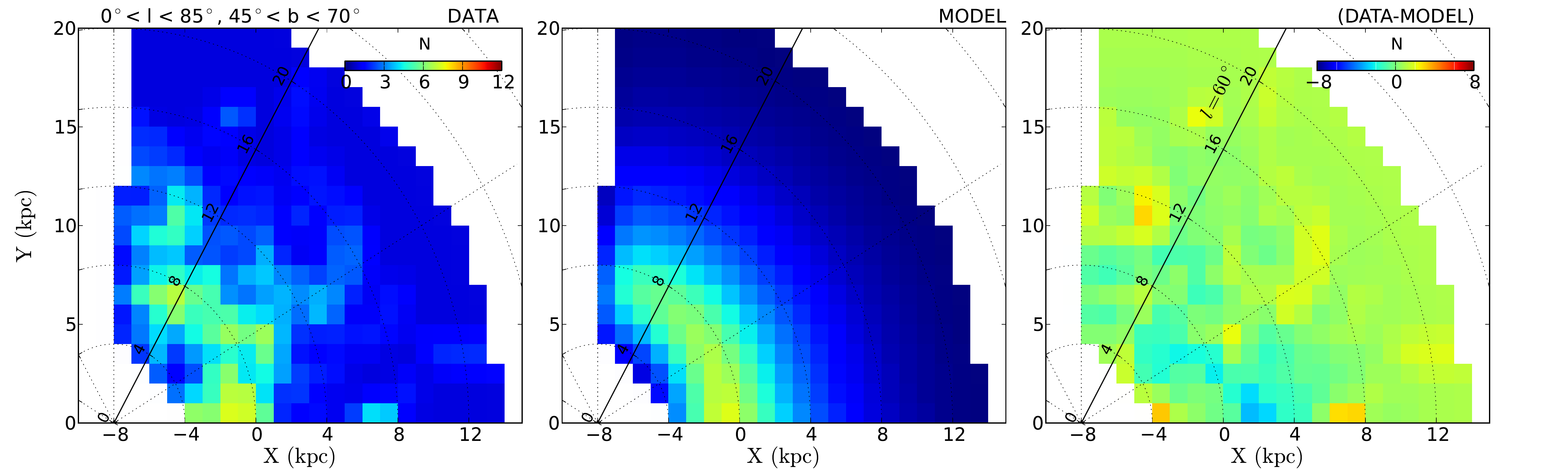}
 \includegraphics[width=155mm]{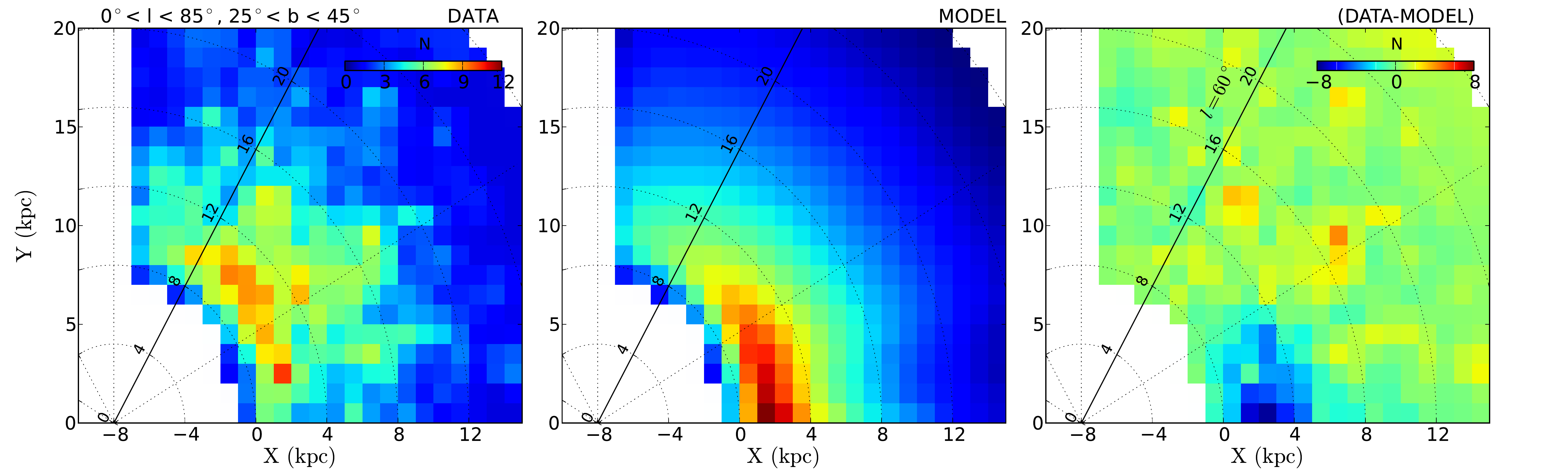}
\includegraphics[width=155mm]{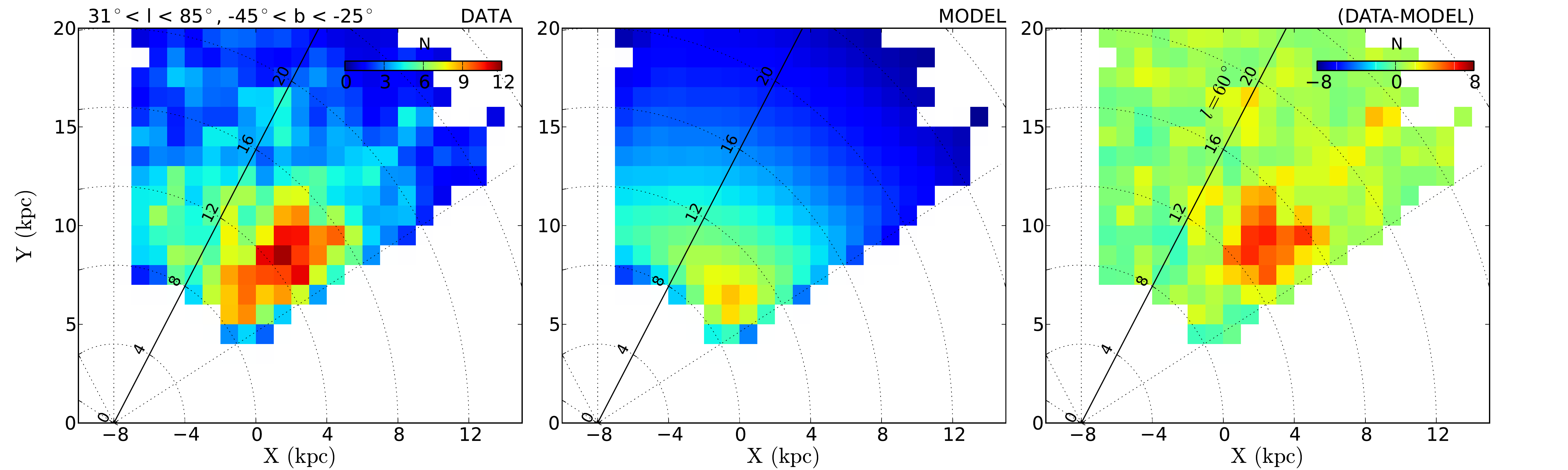}
 \caption{Top panels: X-Y number density maps over a 1132 deg$^{2}$
   field for the data (left) and the model (middle) in the
   control-field outlined in blue in Figure~\ref{footprints}. These
   maps indicate the good agreement between the data and the model in
   this region of the sky where no known halo substructure is present
   in the distance range considered.  Middle panels: X-Y maps over a
   1385 deg$^{2}$ field that includes our selection of the NH
   and NH2 fields. We see a small overdensity at $30^{\circ} < l <
   50^{\circ}$ but also a strong under density for $l < 20^{\circ}$,
   which is due to the incompleteness of the survey in a region
   of high extinction, as shown by the black contours of constant
   E(B-V) reddening in Figure~\ref{lbmap}.  Lower panels: X-Y map over
   a 880 deg$^{2}$ region which includes the SH ans SH2 fields. The residuals
   reveal the RRL excess in the SH at projected distances on the
   Galactic plane between 12 and 16 kpc. The maps have been smoothed with    with a $\sigma = 0.6$ Gaussian filter and are limited to the $10<D<30$ kpc distance range. The origin marks the location of the Galactic center (the Sun is at $X = -8$; $Y = 0$ kpc). }
 \label{xymaps}
\end{figure*} 
For the BPL model of \citet{Deason2011} derived using BHB stars, the
RR Lyrae density normalisation is obtained using the following
procedure. The model of \citet{Deason2011} and the CSS RRab data are
both integrated over the area covered by the control field shown in
blue in Figure 1. The control field is a 1,100 deg$^2$ region with
$0^{\circ} < l < 80^{\circ}$ and $45^{\circ} < b < 70^{\circ}$. To our
knowledge, there are no stellar streams or satellites reported in this
area within distances D$_{RR} <$ 30 kpc. Equating the
incompleteness-corrected integral of the BPL model of
\citet{Deason2011} over the control field area out to $D=30$ kpc to
the total number of RRab in the CSS data we obtain $\rho_{\odot}^{RR}
= 7.3$ kpc$^{-3}$, shown in italics in Table~\ref{Table1}. 

Figure~\ref{control-field} compares the CSS RRab
data in the control field and the predictions of the five simple
power-law models listed in Table~\ref{Table1}. The data is shown in the blue
histograms using 2 kpc binning (errors on the distance are $\sim$ 7\%, i.e. less
than the bin size).  The uncertainty in the number of expected RR
Lyrae is calculated as $\delta N_{obs} \approx \sqrt{(\delta N_{exp})^{2} + (\delta F)^{2}}$, where $\delta F$ is the uncertainty in the completeness
estimated in each distance bin and $\delta N_{exp} $ is the Poisson
uncertainty $\sqrt{N_{exp}}$. According to the Figure, single power-law
models can not reproduce the data adequately. In order not to
over-estimate the number of tracers at large distances, SPL models
must under-predict the counts at smaller distances: note how both red
dashed and dash-dotted curves fall low in the distance range $5 < D <
15$ kpc. Of the three broken power-law models, the ones by
\citet{Deason2011} and \citet{Sesar2011} perform the best with the
model of \citet{Watkins2009} consistently over-predicting RR Lyrae
counts at $D>10$ kpc, albeit at the level $<10\%$ \footnote{Note that
  the \citep{Watkins2009} model has to be scaled down by a factor of
  11 in agreement with DR13b.}. In what follows, we choose to compare
the data with the model of \citet{Deason2011} simply because it gives
marginally higher RRab counts at $D>10$ kpc, and therefore compared to
the prediction of \citet{Sesar2011} the estimate of sub-structure
excess is slightly less optimistic.


%
\begin{figure*}
\includegraphics[width=56mm]{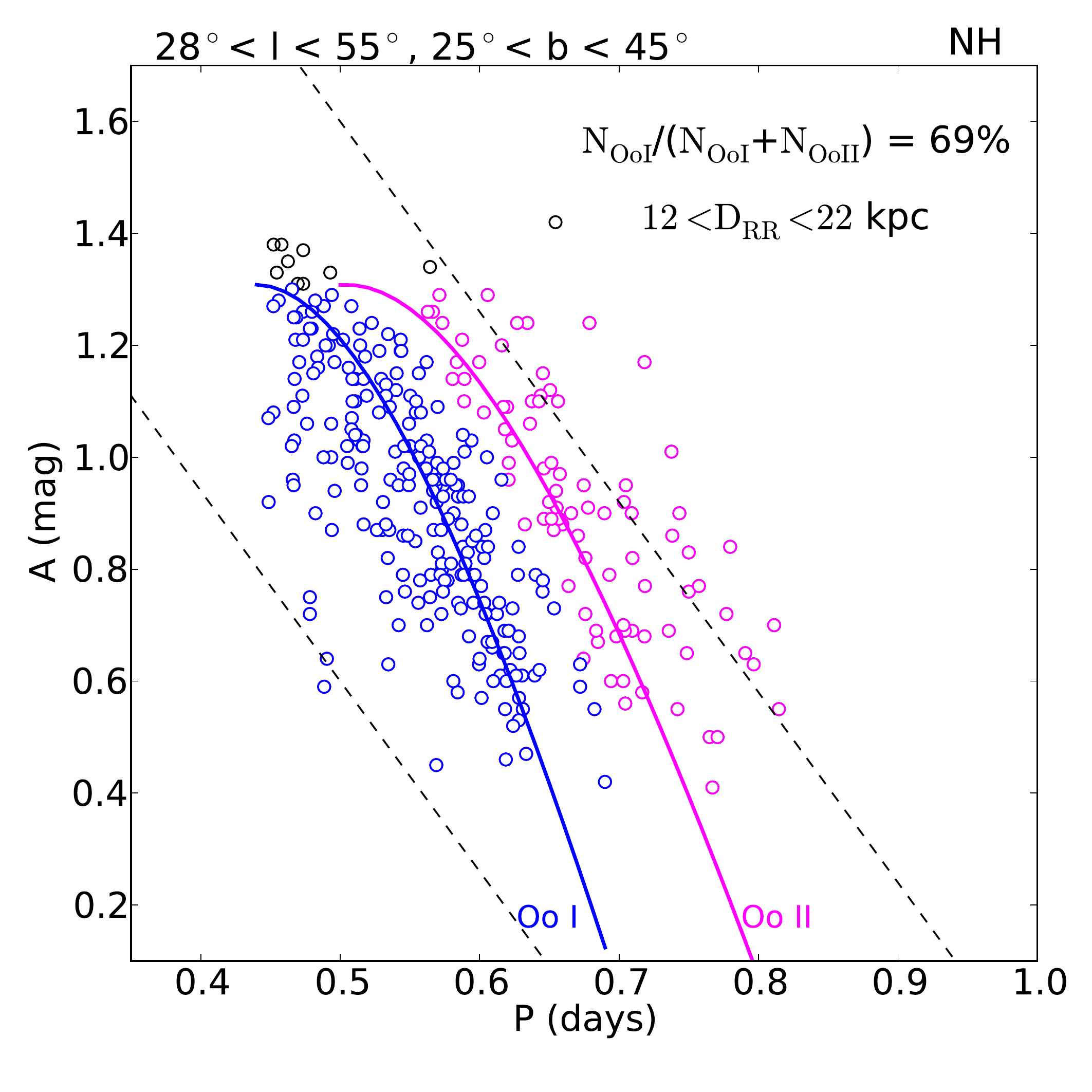}
\includegraphics[width=56mm]{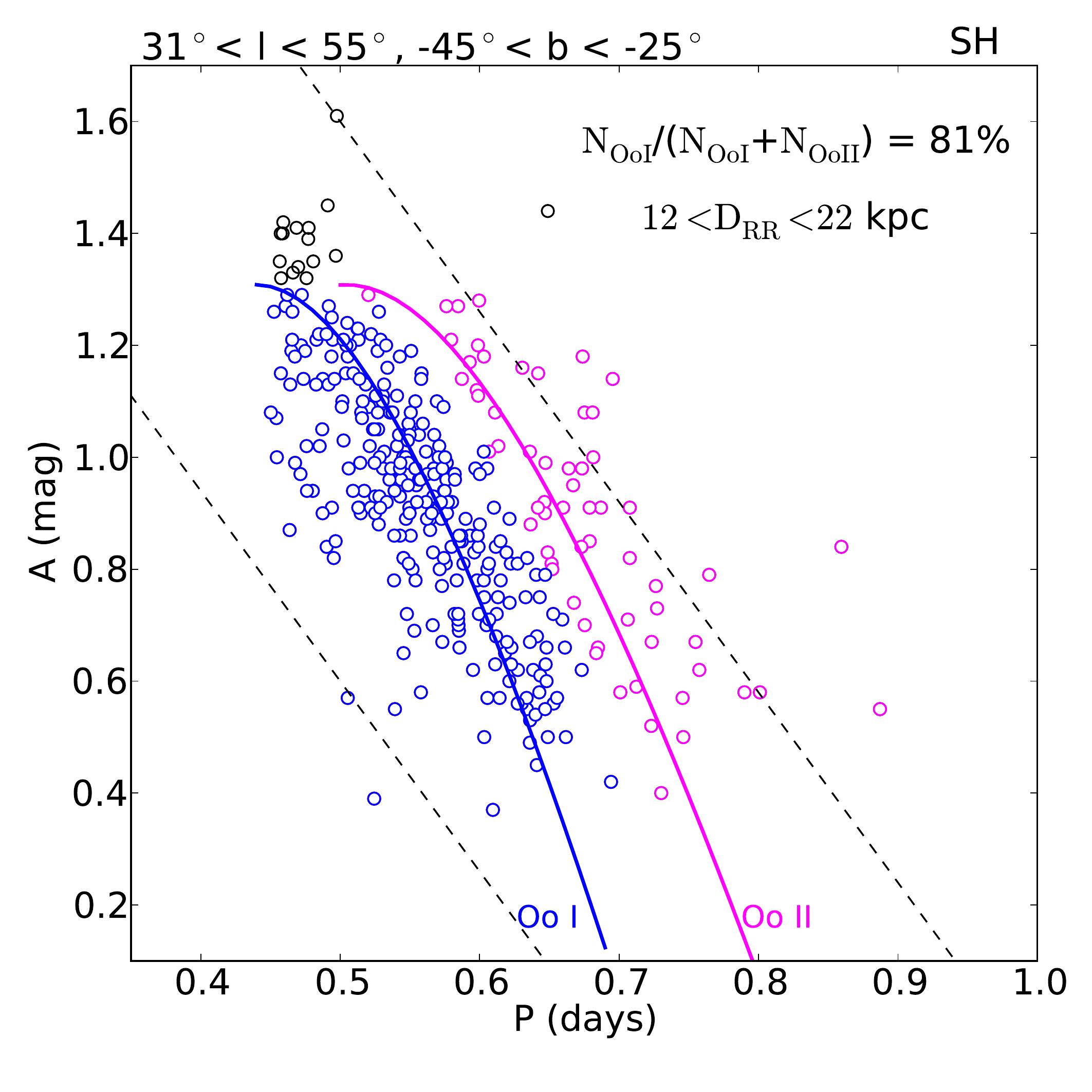}\\
\vspace{-0.3cm}
\includegraphics[width=64mm]{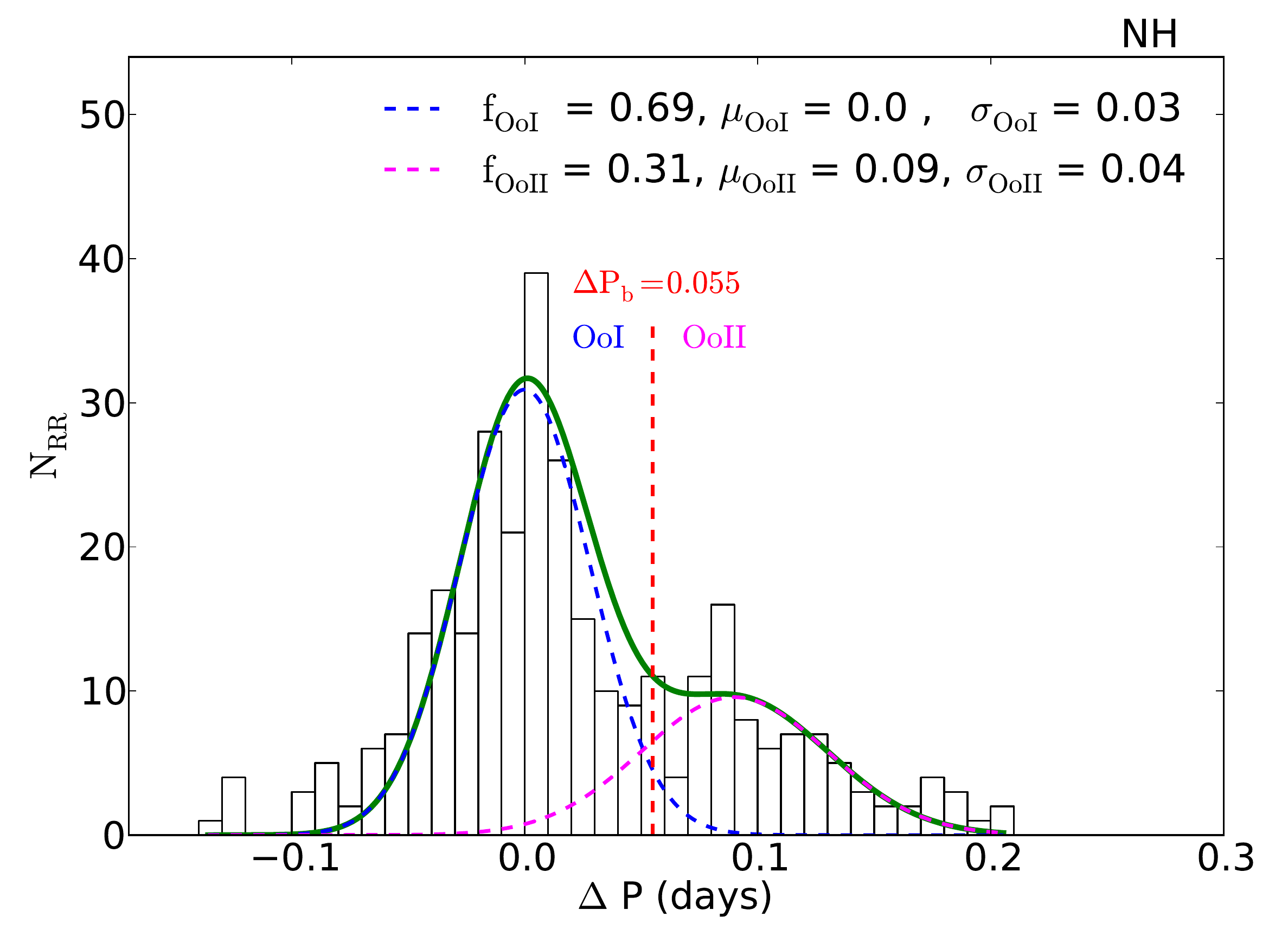}
\includegraphics[width=64mm]{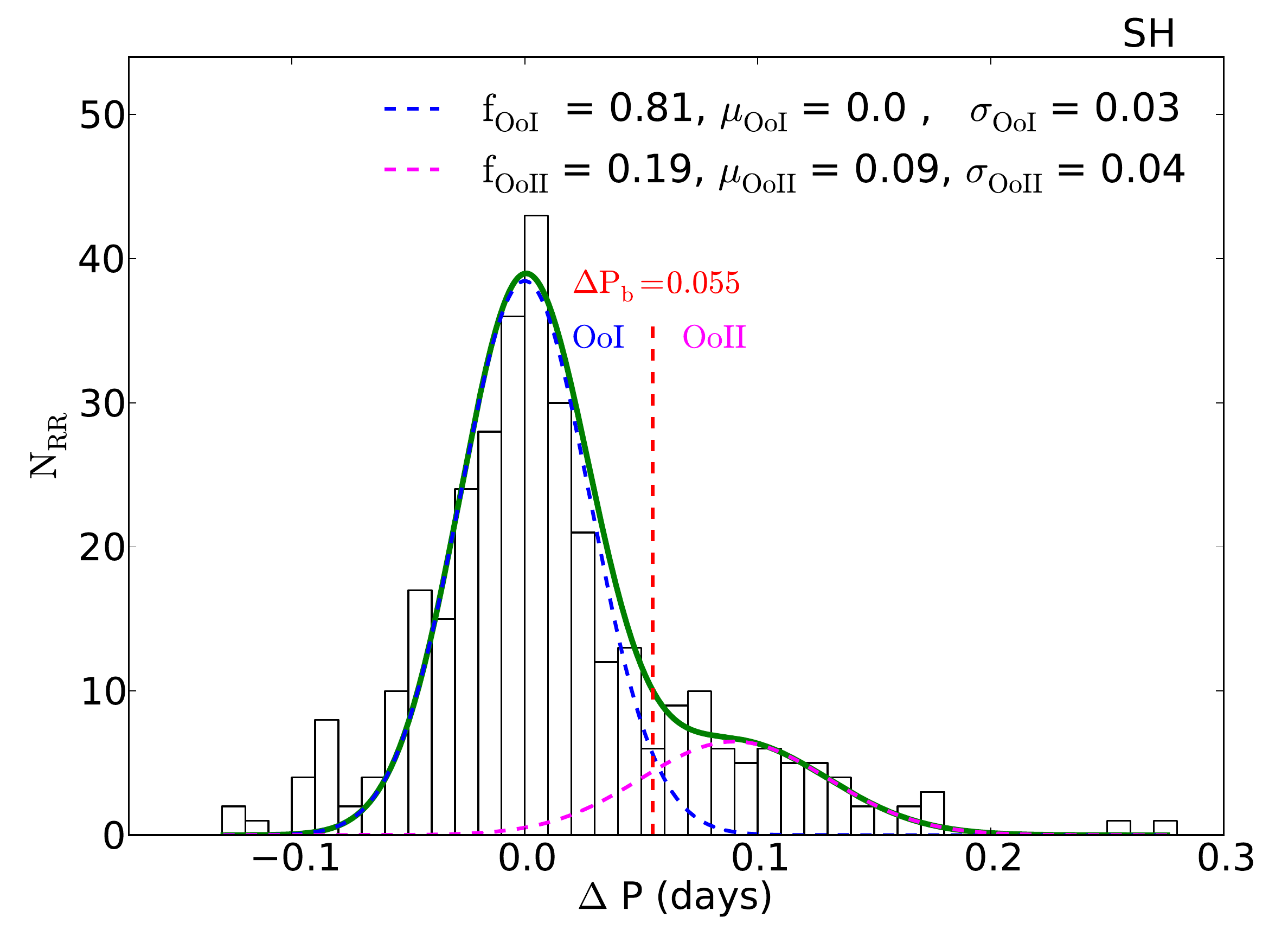}
\caption{Top panels: Bailey diagrams for the RRL in the NH and SH
  fields, at $12<D<22$ kpc. The continuous lines indicate the loci for
  the Oo I and Oo II components as defined by
  \citet{Zorotovic2010}. Blue and magenta indicate the Oo I and Oo II
  RRL respectively, tentatively classified using the $\Delta P =
  0.055$ boundary. The dashed black lines represent the lower $(A =
  2.3-3.4 P ) $ and upper $(A = 3.3-3.4 P )$ limits expected for most
  of the RRab. Bottom panels: Histogram of the period shifts
  calculated for each RRL at constant amplitude from the Oo I locus
  (blue line in the top panels) in the period-amplitude plane defined
  by \citet{Zorotovic2010}. In the NH field 69\% of the stars are Oo I
  type while in the SH, 81\% of them. The expected proportion of Oo I
  RRab is $\sim$75\%. See the text for more details.}
 \label{NH_Oo}
\end{figure*} 

\section{RR Lyrae in the Hercules-Aquila Cloud}
Figure~\ref{lbmap} gives the all-sky density distribution of RR Lyrae
with $10 < D < 25$ kpc in the CSS and in the model of
\citet{Deason2011} applying the re-normalisation deduced in the
previous section. The left (data) and the middle (model) columns of
the Figure look very similar, and as illustrated in the right panel
(residuals), the agreement is only broken in the few regions dominated
by the known halo sub-structures. More precisely, at $-100^{\circ} < l
< -50^{\circ}$ and $b > 30^{\circ}$, the Virgo Cloud, i.e. the nearer
portion of it, the VSS \citep[see e.g.][]{Zinn2014} can be clearly
seen connecting to the leading tail of the Sgr stream.  In-between
these two large structures, at approximately constant Galactic
latitude (at the resolution of the map) $b\sim 50^{\circ}$ runs the
narrow Orphan Stream. The Sgr trailing tail is under the disk, at
$50^{\circ} < l < 180^{\circ}$ and $b < -60^{\circ}$. Comparing the
RRab density in the Galactic North and South at $30^{\circ} < l <
50^{\circ}$, an excess at $-50^{\circ} < b < -20^{\circ}$ is obvious,
roughly bounded by the SH on-cloud field. While the top row gives the
distribution of RRab at all distances over the whole sky, the second,
third and fourth row of the Figure zoom-in on the central
$180^{\circ}$ in the Galaxy and map the RRab in three distance bins:
$10 < D < 15$ kpc, $15 < D < 20$ kpc and $20 < D < 25$ kpc. Judging by
the evolution of the stellar density in the residual map inside the SH
on-cloud field, the peak of the HAC is clearly located in the $15 <
D_{RR} < 20$ kpc distance range, as also shown in
Figure~\ref{RRL-HAC}. In the North, the RR Lyrae overdensity strength
in the on-cloud field is greatly reduced as compared to the South. It
is quite possible that the HAC is asymmetric with respect to the plane
of the disk. However, there could be a more prosaic explanation for
such an observation. Grey contours in the Figure~\ref{lbmap} reveal
the levels of the dust extinction in the Galaxy as deduced by
\citet{Schlegel1998}. According to this map, the reddening is patchy
on rather large angular scales. In particular, a spur of dust reaches
up to higher Galactic latitudes in the Northern hemisphere. Therefore,
this extra extinction could be responsible for diminished star counts
in the NH field. Clearly, the extinction by dust plays some role in
shaping the apparent stellar halo density at lower Galactic $l$ and
$b$ as illustrated in the top right panel of the Figure, where an {\it
  under-density} is observed directly above the bulge, falling neatly
inside the reddening contours.

According to previous studies \citep[e.g.][]{Be2007,Watkins2009},
the HAC is situated in the range $20^{\circ}<l<80^{\circ}$, matching
closely our detection in the Galactic South. To investigate in detail
the line-of-sight distribution of RRab stars in the Cloud, we focus on
two fields in the inner Galaxy, one in the Northern Hemisphere (NH:
$28^{\circ}<l<55^{\circ}$, $25^{\circ}< b < 45^{\circ}$) and one in
the Southern Hemisphere (SH: $31^{\circ}< l < 55^{\circ}$,
$-45^{\circ}< b < -25^{\circ}$). These fields are marked in red in
Figure~\ref{footprints}. Figure~\ref{RRL-HAC} shows the heliocentric
distance number density distribution of RRab in the NH (top) and SH
(bottom) fields together with the model predictions in the left
column, and the relative residuals between the data and the model in
the right column. The errors on the predicted number of RR Lyrae
$\delta N_{obs}$ as a function of distance have been estimated as in
the previous section, and take into account the uncertainty in the
completeness $\delta F$ and in the expected number of counts $\delta
N_{exp}$. As evidenced in the top left panel, overall the RRab
distribution is in reasonable agreement with the model, although it
exceeds the predicted counts slightly, especially at $15 < D < 20$
kpc. As the lower panel of Figure~\ref{RRL-HAC} illustrates, in the
Southern on-cloud field, SH, there exists a strong over-abundance of
RRab stars. The data exceeds the model at all distances beyond 10 kpc,
however the observed distribution shows a clear peak in the 17-19 kpc
distance bin.

To further investigate the extent of the Cloud, Figure~\ref{xymaps}
presents the Galactic plane projections of the density of stars with
$10<D<30$ kpc in the three regions containing the control field
($0^{\circ} < l < 85^{\circ}$, $45^{\circ} < b < 70^{\circ}$, top
row), the NH and NH2 fields ($0^{\circ} < l < 85^{\circ}$, $25^{\circ}
< b < 45^{\circ}$, middle row), as well as the SH and SH2 fields
($31^{\circ} < l < 85^{\circ}$, $-45^{\circ} < b < -25^{\circ}$,
bottom row). The data is shown in the Left, the model predictions in
the Middle and the residuals in the Right panels. The top panels
confirm that the \cite{Deason2011} model is able to reproduce the
broad-brush features of the data in the control-field, as also seen in
Figure~\ref{control-field}. In the middle panels, an under-density is
visible at low Galactic latitudes at $l <30{^\circ}$ (most likely
caused by the unaccounted dust extinction) as described above. However
these also reveal a small excess for $28^{\circ}< l <55^{\circ}$ and
$R > 12$ kpc. Finally, the most dramatic over-density of RRab stars is
located below the Galactic plane, at the projected heliocentric
distances between 12 and 16 kpc, for $30^{\circ}<l<50^{\circ}$ as
evident from the maps shown in the bottom panels.

\subsection{Oosterhoff dichotomy in the North and the South fields}
\begin{figure*}
\vspace{-0.3cm} \includegraphics[width=44mm]{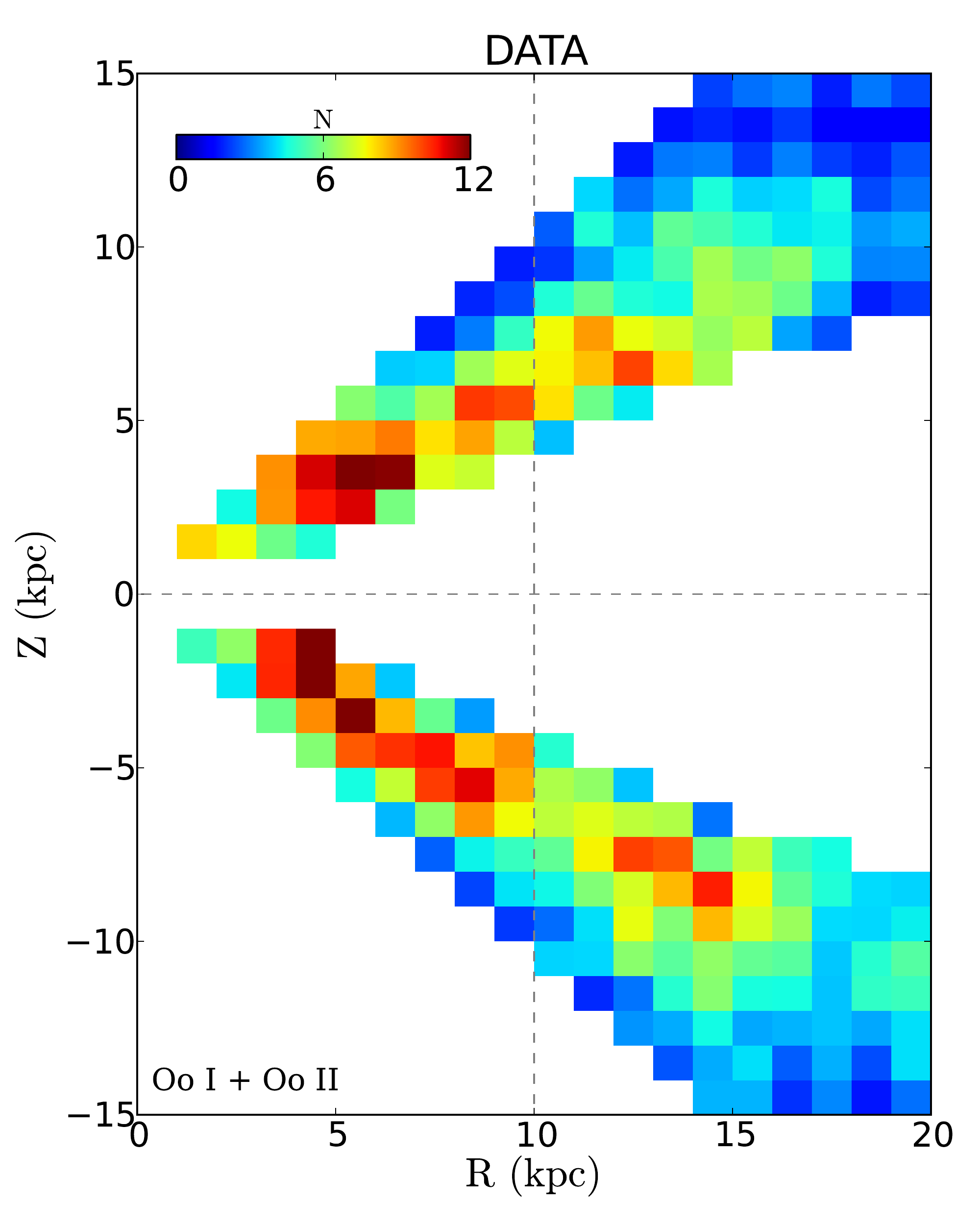}
\includegraphics[width=44mm]{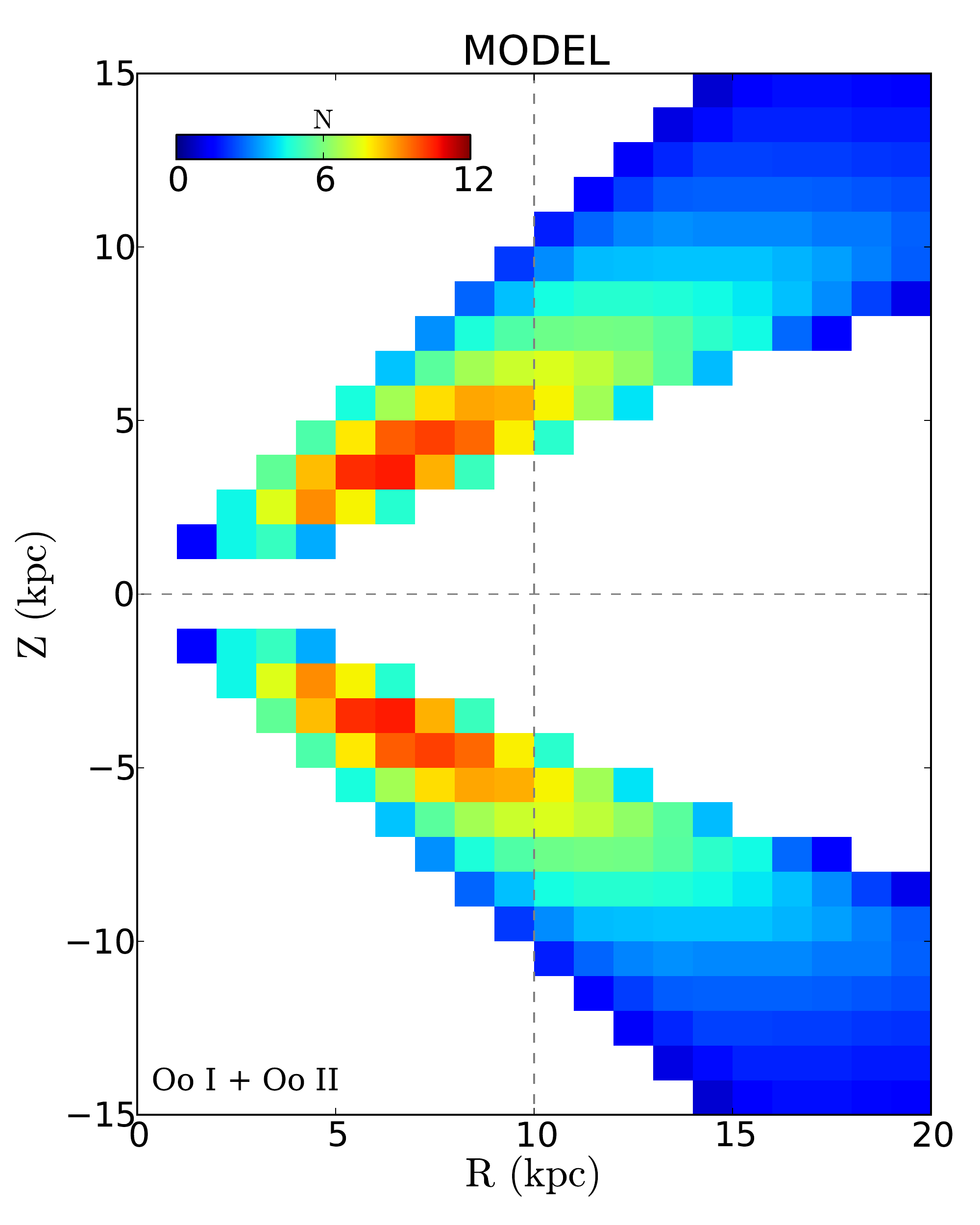}
\includegraphics[width=44mm]{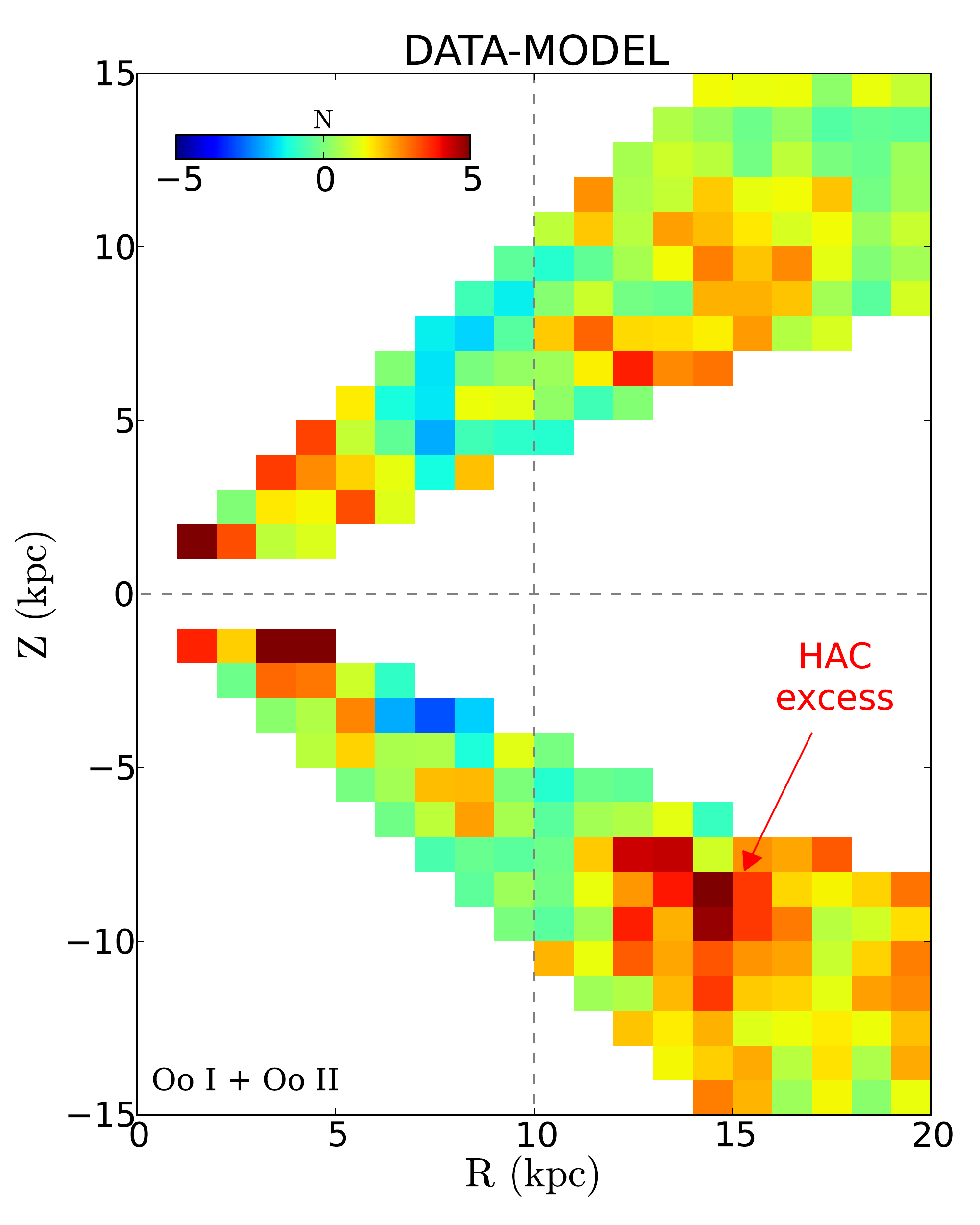}\\ \includegraphics[width=44mm]{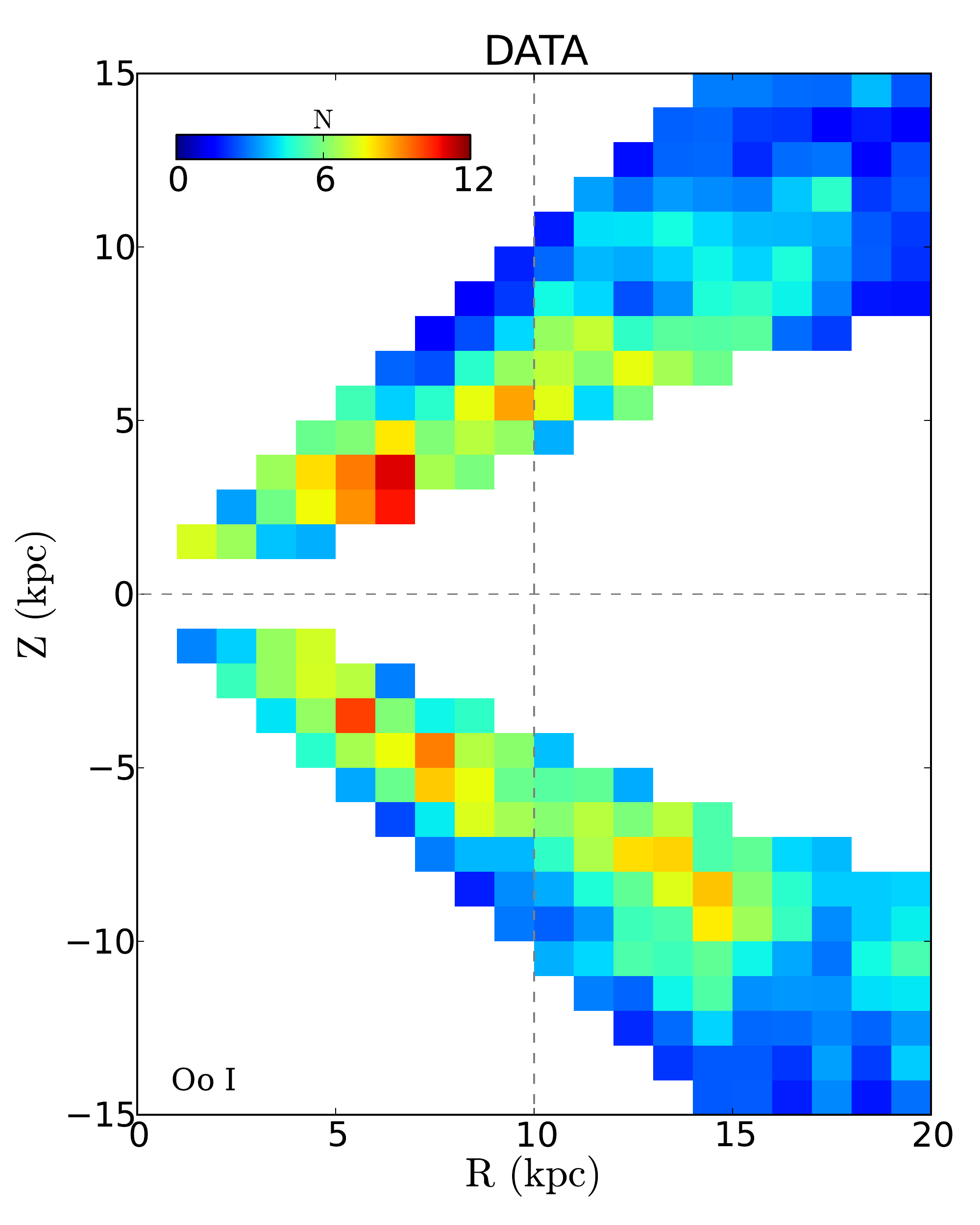}
\includegraphics[width=44mm]{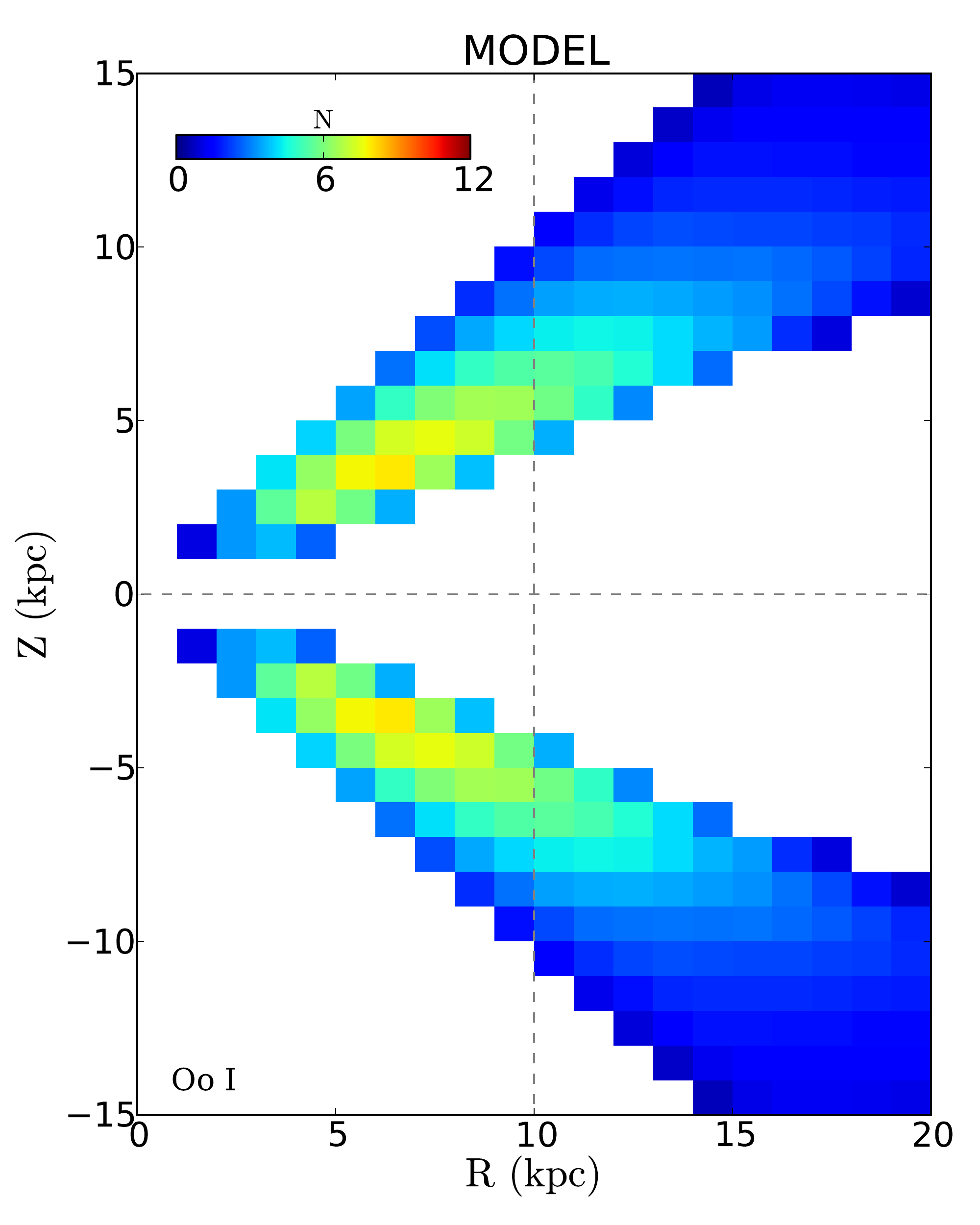}
\includegraphics[width=44mm]{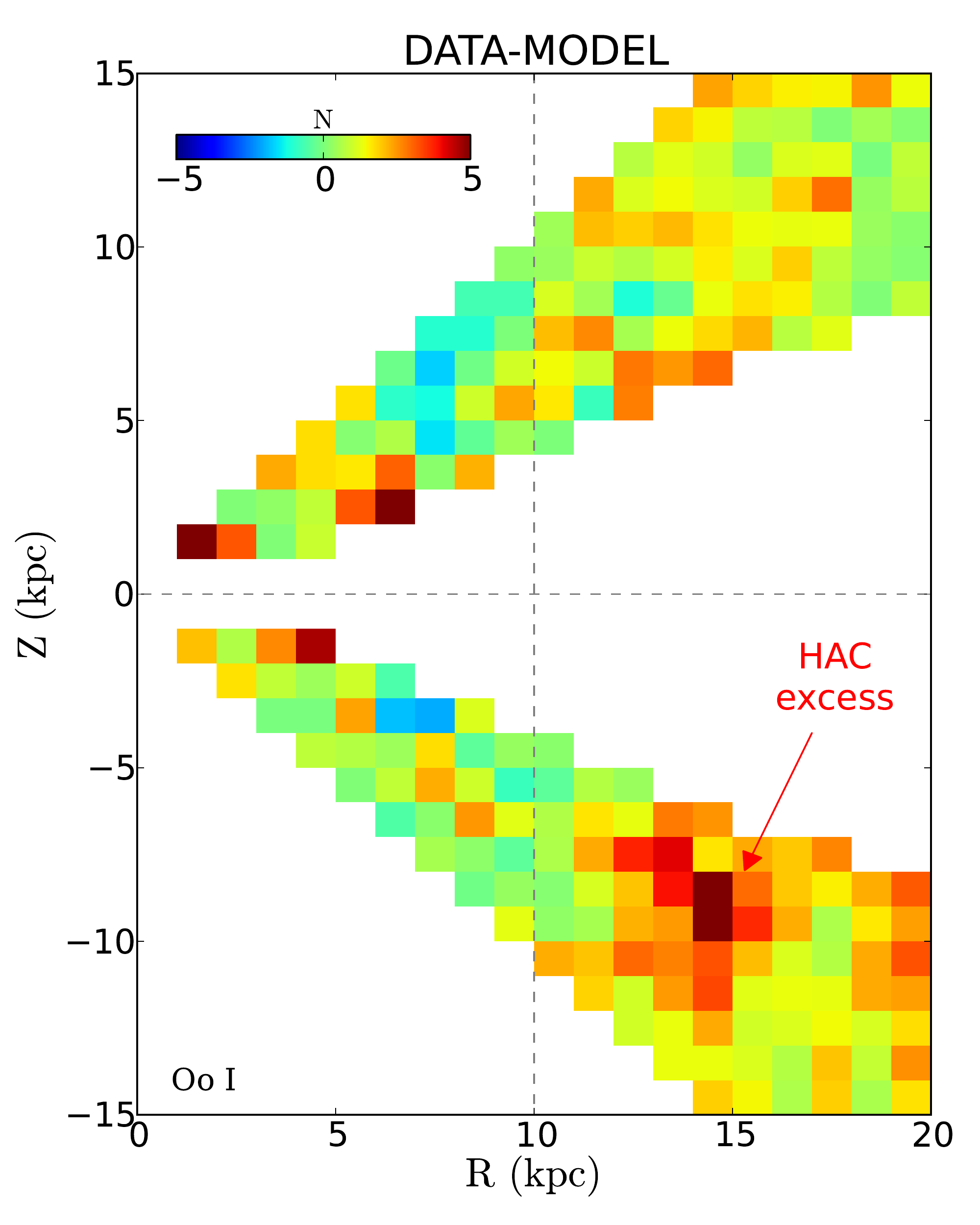}
\includegraphics[width=44mm]{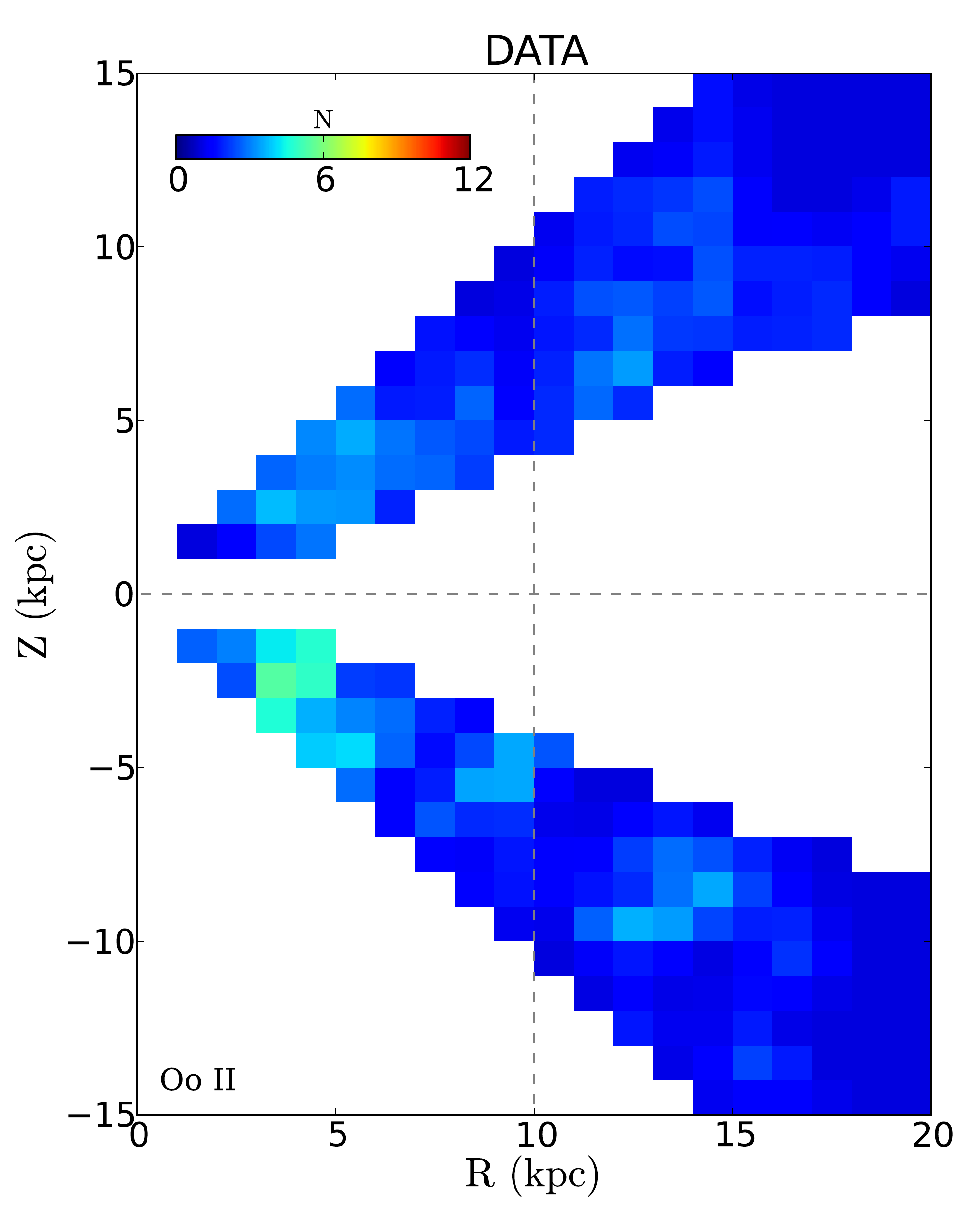}
\includegraphics[width=44mm]{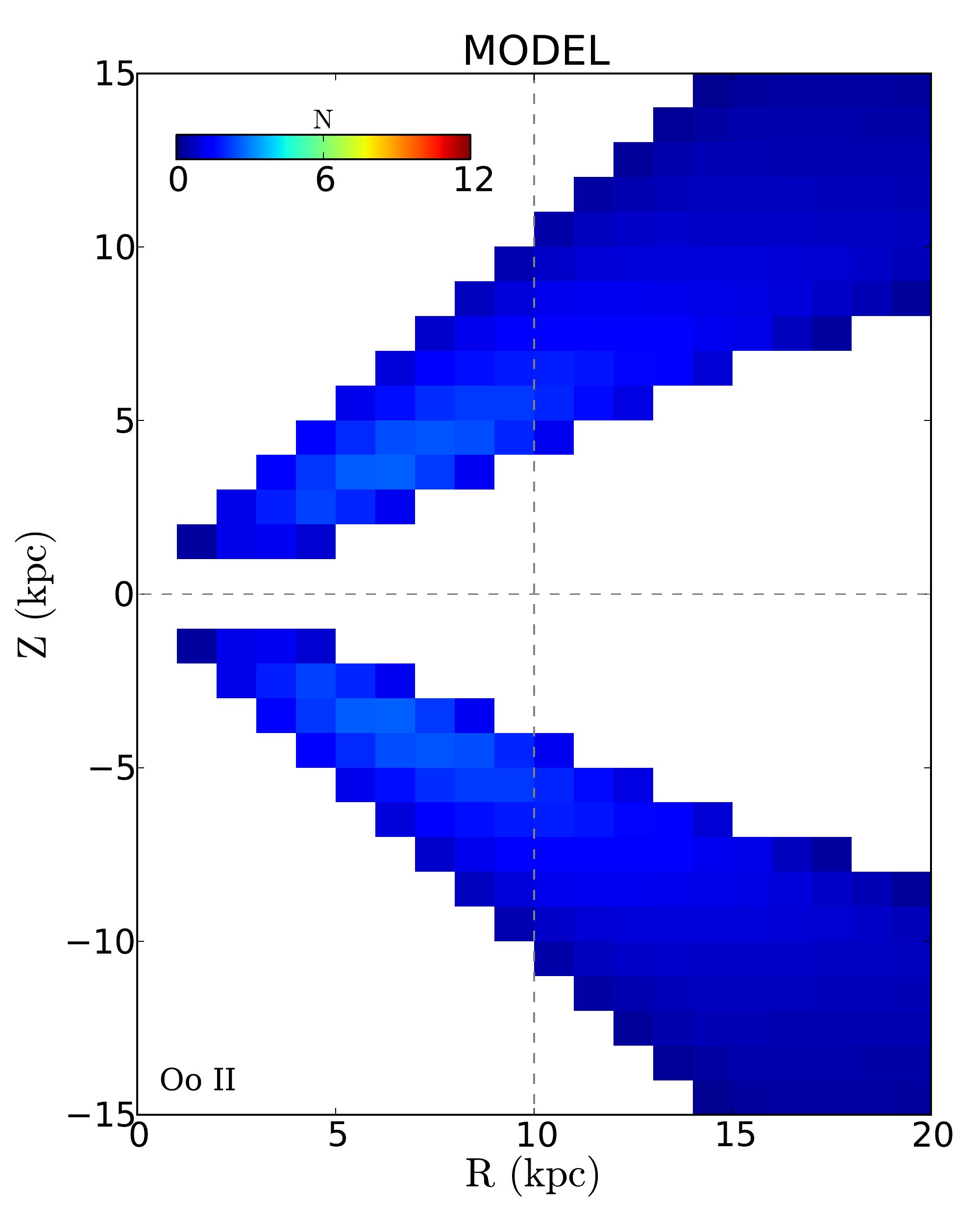}
\includegraphics[width=44mm]{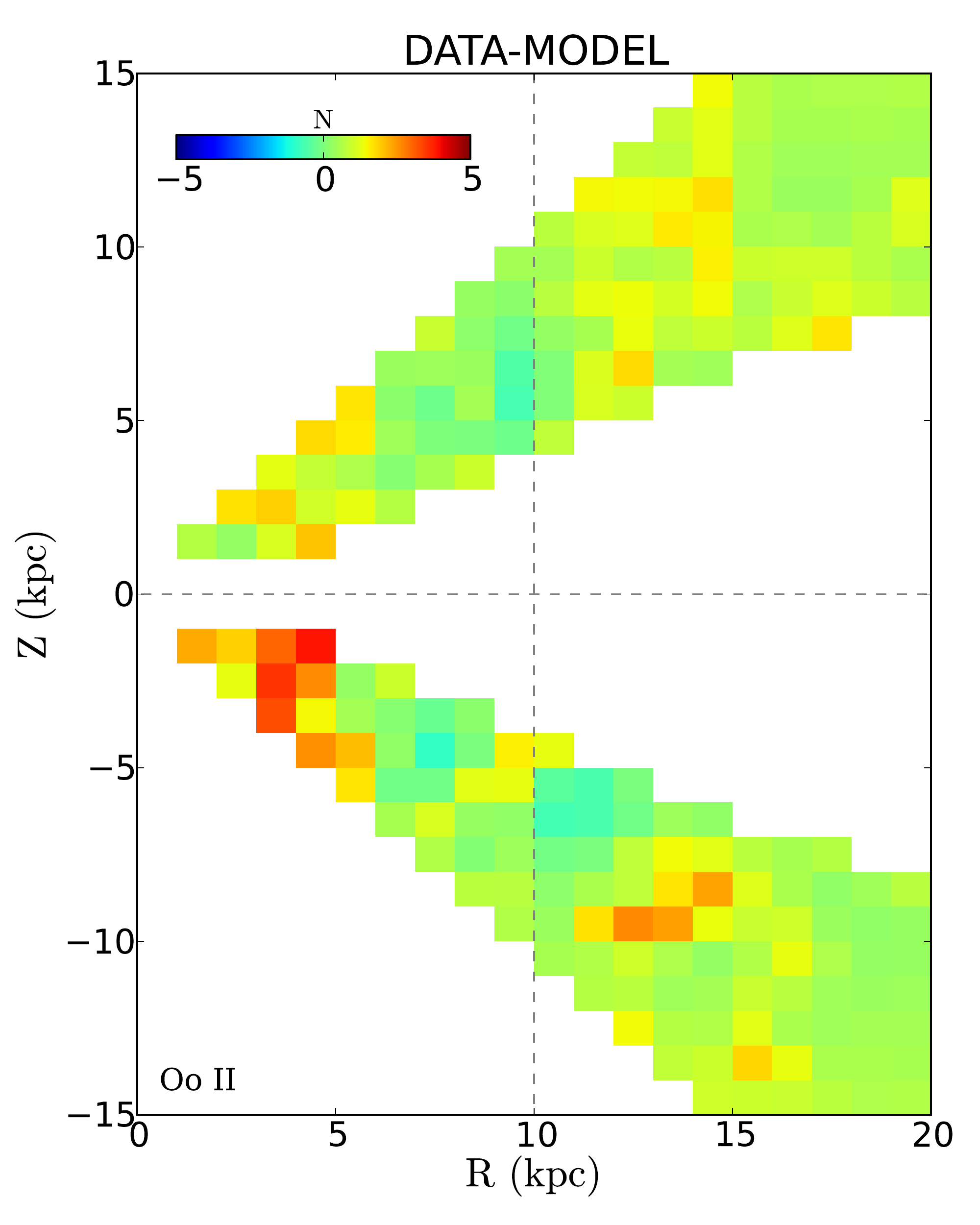}
\caption{R vs Z number density maps of all the RRab Lyrae (top row),
  Oo I RRL (middle row) and Oo II RRL (bottom row) for the data (left
  column) and the \citet{Deason2011} model (middle column) in the N-S
  symmetric fields $31^{\circ}<l<55^{\circ}$, $|b|<45^{\circ}$.The
  residual maps (right column) reveal an overdensity of Oo I RRab in
  the Southern Hemisphere, for $R > 10$ kpc. A much weaker signal is
  present also in the Northern Hemisphere for $R > 12$ kpc, as already
  seen in Figure~\ref{xymaps}. The model calculations take into
  account the completeness function of the survey. The bin size is 0.5
  kpc x 0.5 kpc. The statistical significances of the overdense
  regions are listed in Table~\ref{statistics}. The plots have been
  smoothed with a Gaussian kernel of $\sigma = 0.6$.}
 \label{Oo}
\end{figure*}  

The RR Lyrae period-amplitude plane is not populated uniformly.  There
appears to be significant clumping in the distribution, also known as
the Oosterhoff dichotomy \citep{Oo1939}. This bimodality, most
pronounced for members of the Galactic star clusters, is suspected to
originate mostly from the metallicity differences amongst the
population \citep[see e.g. a review by][]{Catelan2009}. The Oosterhoff
I (Oo I) type globular clusters ($<P_{ab}> \sim 0.55$ days) are more
metal-rich compared to the Oosterhoff II (Oo II) globular clusters
($<P_{ab}> \sim 0.64$ days). One hypothesis is that Oo II type
clusters formed early in the proto-Galaxy while Oo I type clusters
might have formed some 2 to 3 Gyrs later
\citep[e.g.][]{Lee1999}. However, GC ages measured with the help of
HST ACS photometry \citep[e.g.][]{Dotter2010} reveal that the age
differences might be substantially smaller. Many of the neighbouring
dwarf spheroidal galaxies and their globular clusters do not display
the Oosterhoff dichotomy but fall instead in the so-called Oosterhoff
gap on the period-amplitude (Bailey) diagram with $0.58<P_{ab}<0.62$
days. It is therefore clear that the distribution of lightcurve
properties of the Galactic RR Lyrae contains complementary information
about the accretion history of the stellar halo itself harboring a
larger fraction of Oo I type stars \citep[e.g.][]{Catelan2008}.

The two top panels of Figure~\ref{NH_Oo} show the so-called Bailey
diagrams for CSS RRab with heliocentric distances between 12 and 22
kpc in the NH and SH fields (c.f. Figure~\ref{lbmap} in the previous
section). We trace the loci for Oo I and Oo II type globular clusters
(blue and magenta lines respectively) using the period-amplitude
relation defined by \citet{Zorotovic2010} and determine the period
shifts $\Delta P (A)$ (i.e. the offset in period at constant amplitude
from the Oo I locus line) for each RRab. In the bottom panels of
Figure~\ref{NH_Oo} we show the $\Delta P$ period-shift distribution
for the RRab stars in the NH ( $28^{\circ}<l<55^{\circ}$,
$25^{\circ}<b<45^{\circ}$, left) and in the SH (
$31^{\circ}<l<55^{\circ}$, $-45^{\circ}<b<-25^{\circ}$, right)
fields. The distribution of $\Delta$P values is centered on
$\Delta$P$=$ 0 by definition (the position of the Oo I locus). However
a tail at long periods is noticeable; this is due to the Oo II
component. By fitting the period shift distribution with a
two-component Gaussian model we estimate that in the NH field,
$\sim$ 69\% of type-ab RRL belong to the Oo I component and 31\% to
the Oo II in the distance range considered (left panels,
Figure~\ref{NH_Oo}). In the SH field the proportion is significantly
different, with the majority of RRab (81\%) belonging to Oo I
population and only 19\% to the Oo II population (right panels,
Figure~\ref{NH_Oo}).
\begin{figure*}
\includegraphics[width=82mm]{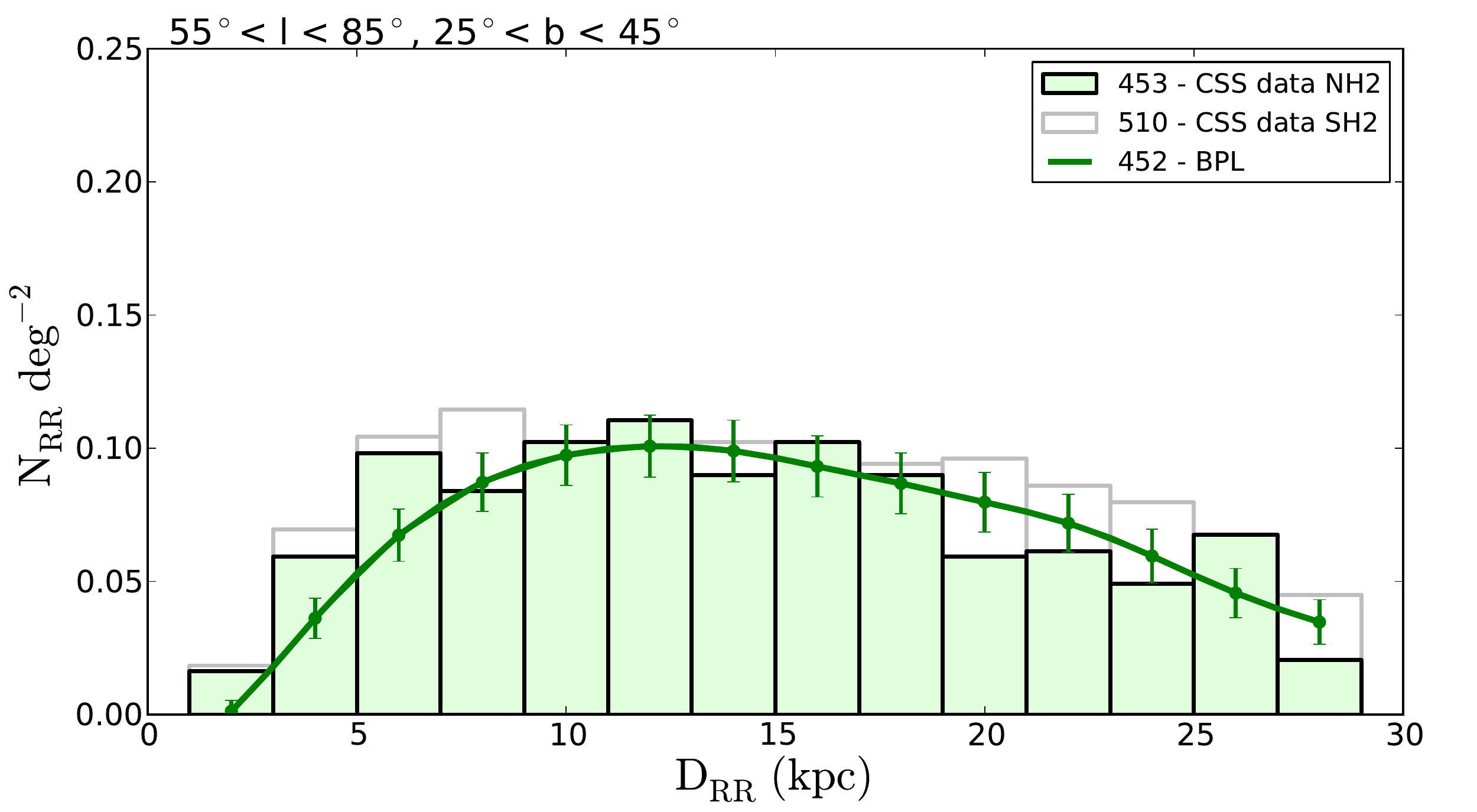}
\includegraphics[width=82mm]{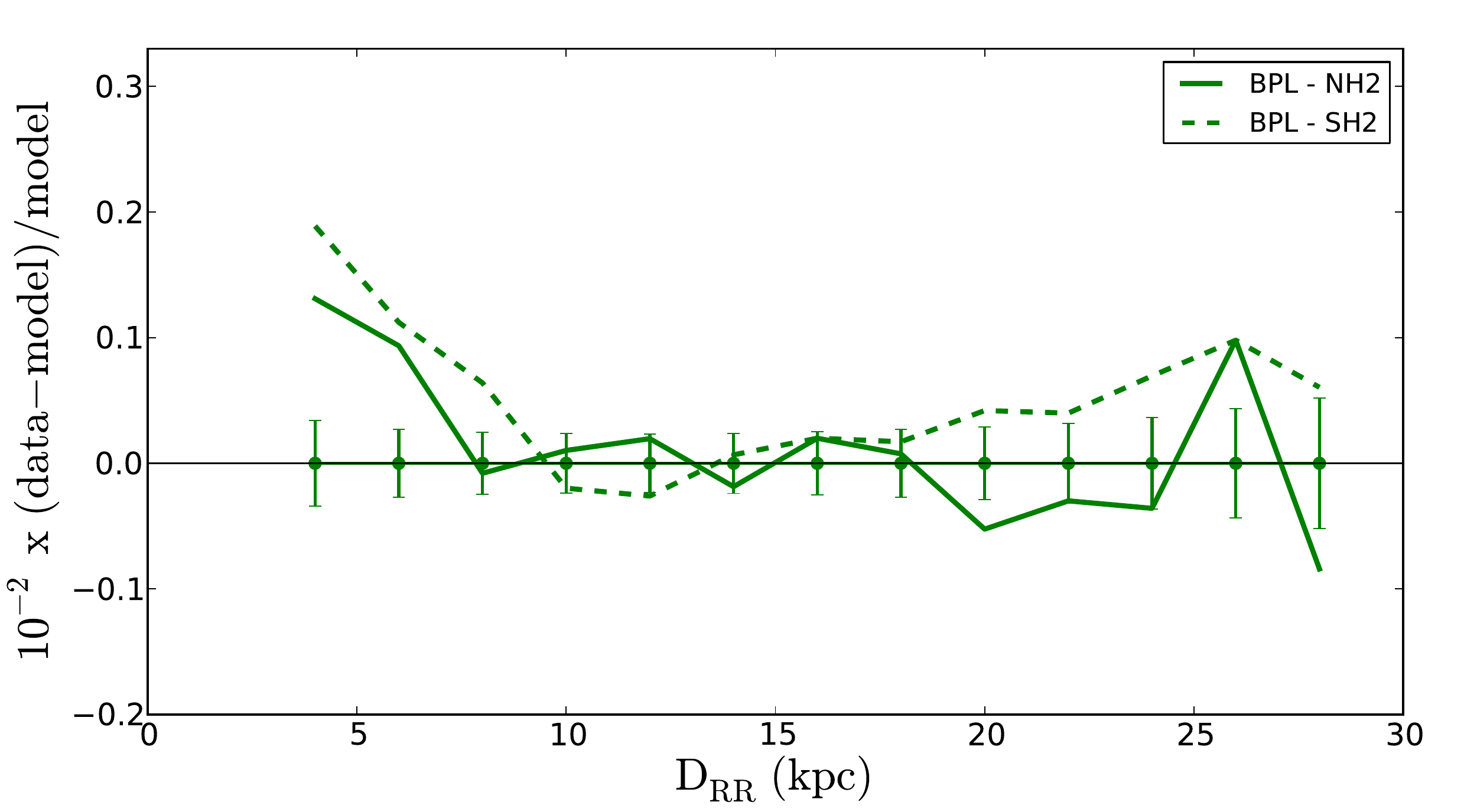}
 \caption{Number density distributions of CSS RRab in the fields NH2 and SH2 adjacent to the NH and SH fields (see Figure~\ref{footprints}). The \citet{Deason2011} model agrees with the observed distribution. The absence of a significant overdensity in these regions suggests that the HAC extends mainly to the NH and (mostly) SH fields.}
 \label{1-2}
\end{figure*} 
To split the sample into Oo I and Oo II populations a simple
boundary at constant period-shift of $\Delta P = 0.055$ (red line in
the bottom panels) is used: Oo I type lie to the left of the boundary and 
Oo II type to the right. In the Bailey diagrams (top panels in
Figure~\ref{NH_Oo}) for the NH and SH fields the Oo I and Oo II types RRL are 
marked with blue and magenta respectively while black indicates the stars 
that failed this simple classification, because situated above the defined loci. Previous studies \citep[e.g.][]{Miceli2008, Sesar2013, Drake2013a} have shown that $\sim 75\%$ of halo
field RRab belong to the Oo I population.  The fact that the SH
field contains $\sim 81 \% $ of these indicates that the progenitor of
the HAC in the Southern Hemisphere is either an Oo I population
globular cluster or a dwarf galaxy falling into the Osterhoff gap. In
fact, using the boundary $\Delta P = 0.055$ we classify most of the
stars in the Oosterhoff gap as Oo I type. Type ab stars that lie to
the left of the Oo I curve may either be metal-rich or have smaller
mean amplitudes because of the Blazhko effect \citep{Bl1907}.
\begin{table}
 \caption{Poisson Statistics for the RRab in the HAC region, shown in
   the top panels of Figure~\ref{Oo}. We list two values for each
   region: RRL with $0 < R < 10$ kpc / $10 < R < 20$ kpc. The 'NH'
   field has $0 < Z < 15$ kpc while the 'SH', $-15 < Z < 0 $ kpc.}
 \label{statistics}
 \begin{tabular}{@{}lcccccc}
 \hline
  RRL Type & Region & $N_{exp}$& $N_{data}$ & Significance \\
 \hline
   Oo I, II& NH& 223/247 & 234/300  & 0.74/3.37\\
   Oo I, II&SH& 223/247 & 241/368  & 1.21/7.70\\
   Oo I& NH& 168/185 & 179/221 & 0.85/2.65 & \\
   \textbf{Oo I} & \textbf{SH}  & \textbf{168/185} & \textbf{168/307} & \textbf{0.00/8.97} &\\
   Oo II& NH& 56/62 & 55/79 & 0.13/2.16& \\
   Oo II& SH  & 56/62 & 73/61 & 2.27/0.37 &\\
\hline
 \end{tabular}
\end{table}

We show the spatial distribution of the Oo I and Oo II
components along the sight-lines towards the HAC
region in Figure~\ref{Oo}.  The density map is shown in the plane of
heliocentric distance $R=\sqrt{X^{2} + Y^{2}}$ and height above the
plane $Z$ for both NH (for this plot, $31^{\circ}<l<55^{\circ}$ instead of $28^{\circ}<l<55^{\circ}$, to allow for symmetry with the SH field) and SH fields.  The left column shows the observed
number density distributions of the RR Lyrae, the middle column gives the
smooth model predictions and finally the right panels report the
resulting residuals. From top to bottom the rows show: the entire RR
Lyrae sample (top); the Oo I type (middle); and the Oo II type
(bottom). The model density normalisations for the individual Oo types
are chosen by simply assuming that Oo I types make up 75\% of the total
RRab population and Oo II the remaining 25\%. The residual maps (right
panels) reveal asymmetries in the North-South distributions. We
quantify the statistical significance of the overdense regions
relative to the model for the (Oo I +Oo II), Oo I and Oo II
populations at heliocentric distances $R$ smaller and greater than 10 kpc. For
each population we compute the statistical significance by finding the
absolute difference between the observed ($N_{data}$) and expected
($N_{exp}$ ) number of RR Lyrae stars, then dividing it by the Poisson
uncertainty ($\sqrt{N_{exp}}$). The results of this analysis are
summarized in Table~\ref{statistics}. We list separately the values
for the regions above (NH) and below (SH) the Galactic plane (see
column 'region') and for each region we compute the significance of
the overdensity for $R < 10$ kpc and $R > 10$ kpc (we list two values
in each column). The values confirm that the model agrees with the
data for $R<10$ kpc, while there is a significant excess of RRL (more
specifically Oo I type) for $R > 10$ kpc. The excess has a statistical
significance of $\sim 9 \sigma$ for the Oo I population and $\sim 8
\sigma$ if we consider the whole sample. While in the SH the
overdensity seems to be due to the Oo I population, in the NH there seems
to be a small excess due to both the Oo I ($\sim 2.6 \sigma$) and Oo
II populations ($\sim 2.2 \sigma$). These values explain the ratios
($\mathrm{f_{OoI}} =$ 69\% for SH and $\mathrm{f_{OoI}} =$ 81\% 
SH) we obtained in Figure~\ref{NH_Oo}. Highlighted in black are
the values for the Oo I component in the SH. Our results are broadly
consistent with those reported in earlier studies. For example, in
Figure 24 in \citet{Sesar2010}, the Hercules–Aquila Cloud is seen at
310$^{\circ} < R.A.< 330^{\circ}$ in the distribution of main-sequence
stars in Stripe 82, at distances in the range 10-25 kpc, with a factor
of $\sim$1.6 overdensity. In the same study, it is found that the HAC
is dominated by Oo I stars (end of section 4.5, \citet{Sesar2010}).

\subsection{Luminosity of the progenitor}

To estimate the total luminosity of the Cloud we need to constrain its spatial 
extent. Figure~\ref{1-2} shows the heliocentric
number density distribution of RRab stars in two regions symmetric
with respect to the Galactic plane, and neighbouring the NH and SH
fields (note boxes marked in green and labelled NH2 and SH2 in Figure~\ref{footprints}). Field NH2 
encompasses stars with $55^{\circ}<l<85^{\circ}$,
$25^{\circ}<b<45^{\circ}$, while field SH2 contains
stars with $55^{\circ}<l<85^{\circ}$, $-45^{\circ}<b<-25^{\circ}$.  These 
distributions show no clear excess.  Therefore, we can conclude that at
the Galactic latitudes probed by the CSS, the HAC members are
predominately located in the NH and SH fields.

Assuming that within the CSS footprint the majority of stars belonging
to the Cloud are concentrated in the NH and SH fields, it is possible
to compute a rough estimate of the initial luminosity of the
progenitor. We count the excess RRab stars in both the NH and SH
fields with respect to the BPL model, in the 10 to 28 kpc distance
range. In the North, this amounts to 76 RRab, and more than double,
i.e. 181 RRab stars in the South. The luminosity of the progenitor is
then estimated for a range of disruption scenarios. For example, the
parent of the Cloud might have been an old and evolved system, with
stellar populations similar to that of a globular cluster, packed with
numerous RR Lyrae. Alternatively, it could have been a system with an
intermediate age population and a smaller proportion of RRL given the
total luminosity. In Table~\ref{progenitor} we list a series of
possible progenitor systems indicating the number of RRab
$N_{RR}^{sys}$ in each one, their luminosity $L_{sys}$ and the
inferred luminosity of the HAC's progenitor $L_{HAC}$, assuming:
\begin{equation}
L_{HAC}/N_{RR}^{HAC} =   L_{sys}/N_{RR}^{sys}
\end{equation}
where $N_{RR}^{HAC} = 367$, counting for a $\sim$70\% completeness of
the survey. The total absolute magnitude of a system, knowing its
luminosity is $M_{V}^{sys} = M_{\odot} - 2.5 log_{10}(
L_{sys}/L_{\odot})$.
\begin{table}
 \caption{Examples of old and intermediate systems and estimates of
   the luminosity of the HAC progenitor. The values in the first three
   columns are from \citet{Clement2001} (updated catalogue of Variable
   Stars in Globular Clusters), \citet{Harris1996, Harris2010} (the 2010 revision of his
   Catalog of Parameters for Milky Way Globular Clusters), and Table 1
   in \citet{Smith2009} (RRL properties of Dwarf Spheroidal
   Galaxies).}
 \label{progenitor}
 \begin{tabular}{@{}lcccccc}
 \hline
   Old pop. & $N_{RR}^{sys}$& $M_{V}^{sys}$ & $ L_{sys}$& $ L_{HAC}$ &$M_{V}^{HAC}$  \\
   GCs & & & ($L_{\odot}$) & ($L_{\odot}$) &   \\
 \hline
   N5272 (M3) & 187 & -8.9 & $3 \cdot 10^{5}$ & $6 \cdot 10^{5}$  & -9.6  \\
    N3201        & 72 & -7.4 & $8.1 \cdot 10^{4}$ &  $4 \cdot 10^{5}$ & -9.2 \\
   N6333(M9)        & 9 & -7.9 &$ 1.3 \cdot 10^{5}$ &$5 \cdot 10^{6}$  & -12 \\
\hline
Interm. pop.  &  &  &   \\
Dwarf galaxies & & &  & &   \\
\hline
Leo I        & 47 & -12.0 &$ 5.4 \cdot10^{6}$ & $ 3 \cdot 10^{7}$ & -14.2 \\
Draco        & 214 & -8.8 &$ 2.8 \cdot10^{5}$ & $ 5 \cdot 10^{5}$ & -9.4 \\
Carina        & 54 & -9.1 &$ 3.7 \cdot10^{5}$ & $3 \cdot 10^{6}$  &  -11.2 \\
Fornax        & 396 & -13.4 & $1.9 \cdot10^{7}$ & $2 \cdot 10^{7}$ & -13.3 \\
  \hline
 \end{tabular}
\end{table}
%
%
%
According to Table~\ref{progenitor}, the luminosity of the progenitor could
be anywhere between $M_{V}^{HAC} \approx -9$ and $M_{V}^{HAC} \approx
-14$. However, we have assumed a constant 70\% completeness at all
distances while the completeness decreases down to 40\% at 30 kpc,
leading us to underestimate the actual number of RRL. It is also
possible that the HAC is not limited to the NH and SH fields: if the
NH and SH overdensities are associated, it is likely that the
overdensity extends at lower galactic latitudes. The total luminosity of the Cloud $M_{V}^{HAC} = -13$ estimated by \citet{Be2007}, is in agreement with
the results found in this section.


\section{Conclusions}
\begin{figure}
\includegraphics[width=80mm]{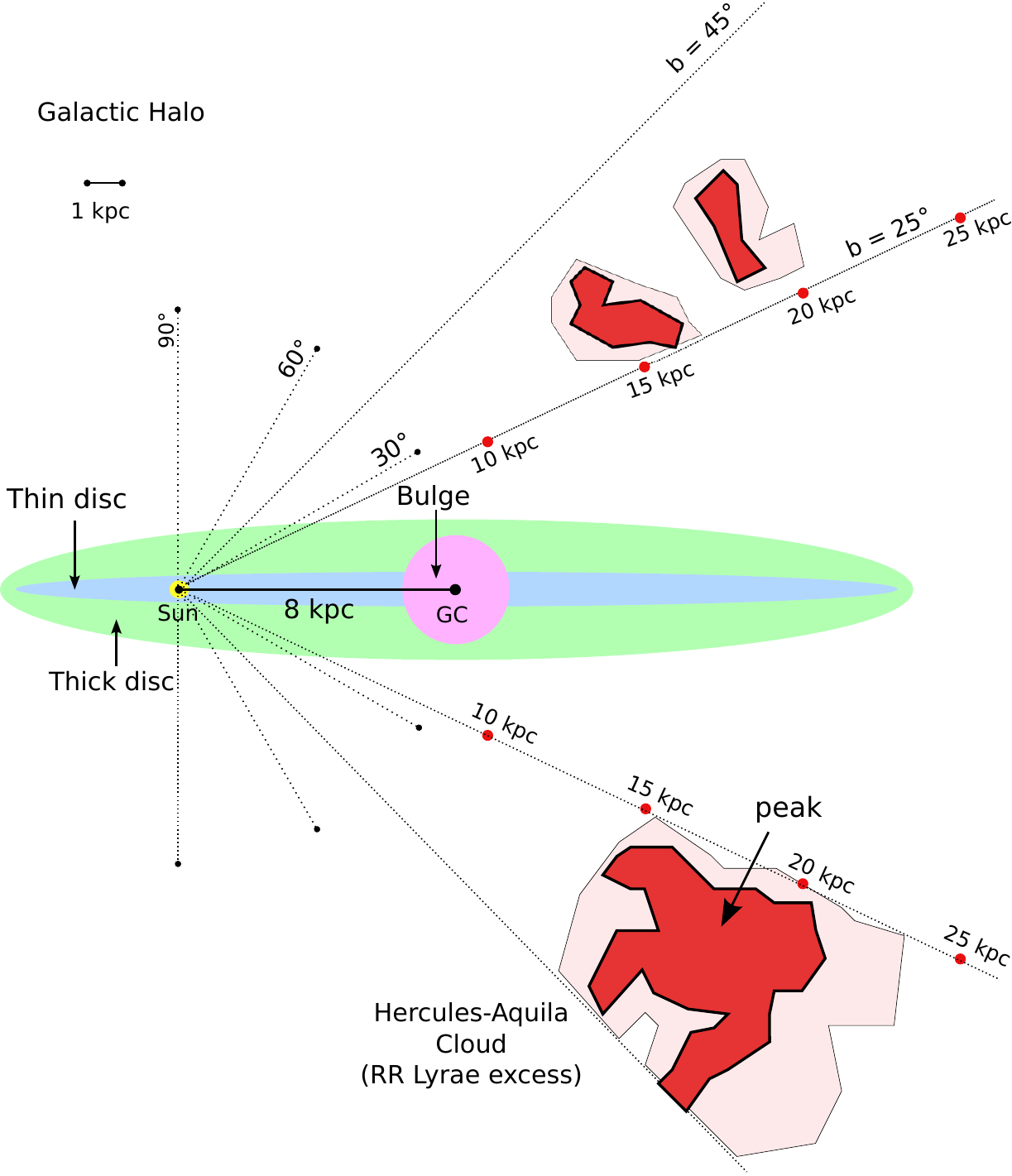}
\caption{Schematic view of the Milky Way structure. We show two
  isodensity contours of the Hercules-Aquila Cloud as revealed by the
  CSS RR Lyrae sample (see also Figure~\ref{Oo}). The peak of the RRL
  over-density is at distances $Z$ between -7 to -13 kpc below the
  Galactic plane, therefore we confidently exclude that the HAC is a
  thick disc structure or orginated from a bar-disc interaction.}
 \label{drawing}
\end{figure} 
We have used a sample of $\sim 14,000$ RR Lyrae from the Catalina
Schmidt Survey to map out the Hercules Aquila Cloud located on the
other side of the Galaxy. To illustrate the location and the extent of
the Cloud, a schematic drawing of the Cloud's signal as traced by the
RR Lyrae is given in Figure~\ref{drawing}. The figure gives a
graphical summary of the results presented in this Paper, which are
also detailed below.

\begin{itemize}

\item In the Galactic Southern hemisphere, there is a prominent overdensity of
  RRab stars in the direction coincident with the previous detections
  of the Hercules-Aquila Overdensity \citep[see
    e.g.][]{Be2007,Watkins2009}. The significance of the RR Lyrae
  overdensity in the South is $\sim 8 \sigma$.

\item In the Galactic Northern hemishpere, the excess is barely noticeable. 
  It is therefore possible that the HAC is not symmetric with respect to the
  disk plane. However, we suspect that the northern signal is greatly
  reduced by the interstellar dust reaching to higher latitudes above
  the disk. Without correcting for this the significance of the RR Lyrae 
  overdensity in the North is $\sim 3 \sigma$.

\item Below the disk, the RR Lyrae excess show a broad peak at $(l,b)
  \sim (40^{\circ}, -30^{\circ})$ (see Figure~\ref{lbmap}), while the
  maximum along the line of sight is reached at a heliocentric of
  $D\sim 18$ kpc. The portion of the Cloud accessible through
  the CSS data is at least 20$^{\circ}$ across. Overall, the physical
  extent of the Cloud as traced by the RRab stars is at least 4 kpc
  along each spatial dimension.

\item The majority of the RR Lyrae stars contributing to the
  overdensity are either of Oo I type or fall into the so-called
  Oosterhoff gap. Therefore, the bulk of the stars in the Cloud
  progenitor seem to resemble the dominant stellar population of
  the Galactic halo.

\item Using the CSS data we have constrained the pre-disruption luminosity 
  of the HAC progenitor to lie in the range $-15 < M_{\rm V} < -9$.

Based on the observations outlined above, we conclude that the HAC is
indeed a distinct stellar halo sub-structure and not part of the
nearby thick disk. Unfortunately, we still lack deep wide-area
infrared data to follow the HAC signal to lower Galactic latitudes,
and therefore our estimate of the Cloud's extent and luminosity remain
rather fuzzy. Fortunately with the identification of the conspicuous RR
Lyrae population in the Cloud reported here, it should now be possible
to map the radial velocity across the face of the Hercules-Aquila
Cloud and thus illuminate the history of its interaction with the Milky
Way.


\section*{Acknowledgements}

We thank Andrew Drake for useful information on the CSS data and the
anonymous referee for the illuminating and thorough review. ITS is
grateful to Adriano Agnello for his constructive comments and
help. This work was partially supported by the Gaia Research for
European Astronomy Training (GREAT-ITN) Marie Curie network, funded
through the European Union Seventh Framework Programme (FP7/2007-2013)
under grant agreement number 264895. The research leading to these
results has also received funding from the European Research Council
under the European Union's Seventh Framework Programme (FP/2007-2013)
/ ERC Grant Agreement n. 308024. VB acknowledges financial support
from the Royal Society.

\end{itemize}

\bibliographystyle{mn2e}
\bibliography{draft8} 
\end{document}